\documentclass[preprint,10pt]{aastex}

\newcommand{\gcc}{\ \mathrm{g\ cm^{-3} }}
\newcommand{\cms}{\ \mathrm{cm \ s^{-1}}}

\newcommand{\eg}{{\emph{e.g.}}}
\newcommand{\ie}{{\emph{i.e.}}}

\newcommand{\imag}{\ensuremath{i}}

\newcommand{\nwd}[1][{}]{\ensuremath{\left( n + W_{#1} D \right ) }}

\newcommand{\magmu}{}
\newcommand{\magH}{B}
\newcommand{\magh}{b}

\newcommand{\coefa}{\ensuremath{C^{(s^-)}}}
\newcommand{\coefb}{\ensuremath{C^{(s^+)}}}
\newcommand{\coefc}{\ensuremath{C^{(e)}}}
\newcommand{\coefd}{\ensuremath{C^{(A^+)}}}
\newcommand{\coefe}{\ensuremath{C^{(A^-)}}}
\newcommand{\coefdp}{\ensuremath{C^{(A^+)'}}}
\newcommand{\coefep}{\ensuremath{C^{(A^-)'}}}

\usepackage{epsf,color}

\begin{document} 

\title{The Linear Instability of Astrophysical Flames in Magnetic Fields}

\shorttitle{Magnetic Fields and Flame Stability}
\shortauthors{L.~J.~Dursi}

%\slugcomment{in preparation, \today}
\slugcomment{Submitted to ApJ, 17 Nov 2003}

\author{ L.~J.~Dursi\altaffilmark{1} }

\altaffiltext{1}{Dept.\ of Astronomy \& Astrophysics, 
                 The University of Chicago, 
                 Chicago, IL  60637}

\begin{abstract}
Supernovae of Type~Ia are used as standard candles for cosmological
observations despite the as yet incomplete understanding of their
explosion mechanism.   In one model, these events are thought to result
from subsonic burning in the core of an accreting Carbon/Oxygen white
dwarf that is accelerated through flame wrinkling and flame instabilities.
Many such white dwarfs have significant magnetic fields.  Here we derive
the linear effects of such magnetic fields on one flame instability,
the well-known Landau-Darrieus instability.  When the magnetic field
is strong enough that the flame is everywhere sub-Alfv\'enic, the
instability can be greatly suppressed.  Super-Alfv\'enic flames are much
less affected by the field, with flames propagating parallel to the field somewhat
destabilized, and flames propagating perpendicular to the field somewhat stabilized.
Trans-Alfv\'enic parallel flames, however, like trans-Alfv\'enic parallel
shocks, are seen to be non-evolutionary; understanding the behavior of
these flames will require careful numerical simulation.  \end{abstract}


% ============================================================================
%  Introduction
% ============================================================================
\section{INTRODUCTION}
\label{sec:intro}
\subsection{Supernovae Type Ia Progenitors}

The current standard model for Type Ia supernova explosions is that a
flame or flames begin inside a Carbon-Oxygen Chandrasekhar mass white
dwarf (see for instance the review by \cite{hillebrandtniemeyer00}).
The burning may accelerate by flame instabilities; the Carbon-Oxygen
flame is unstable to the Landau-Darrieus and Rayleigh-Taylor instabilities
\citep{COld} which wrinkles the flame front, burning a larger surface
area of material per unit time.

Although the class of white dwarfs that forms the progenitors for Type Ia
supernovae is unclear, it is known that many white dwarfs have significant
surface magnetic fields, up to $10^8$~G for accreting systems and isolated
systems with magnetic fields up to $10^9$~G are known; central magnetic
fields may be much larger (see the review by \citealt{magwdreview},
and new data in for instance \citealt{sdssmagwhitedwarfs}).  Fields of this strength
may certainly affect the dynamics of flame instabilities, and thus the
mechanism of supernovae.  It is known that magnetic fields can destabilize
shocks (see for example \citealt{stoneedelman} and references therein),
and suppress other interfacial instabilities, such as Richtmeyer-Meshkov
\citep{rmsupression} and Rayleigh-Taylor \citep{chandra}.

In terrestrial chemical combustion, many intermediate species are ionized, so that
magnetic fields might  be thought to have a dynamic effect.  However,
research on flames in magnetic fields has been sparse in the terrestrial
flame literature because the diffusion of magnetic fields under
in terrestrial burning conditions is fast enough that any interaction
between the flame and the magnetic field is quite limited.  Some recent
work has been done on the effect of ambient magnetic fields on chemical
flames \citep{maglaminarflames,magdiffusionflames} where small effects
were found mainly due to the effect on large-scale motions of paramagnetic
gases such as $O_2$.

In the astrophysical case magnetic fields may be more relevant,
and the flame-field interactions may be richer in this extremely large
magnetic Reynolds number regime, as perturbations in the magnetic
field cannot readily diffuse away.  Although there has been some look at
large-scale burning dynamics in the presence of a large magnetic field
\citep{ghezzi}, little work has been presented on the microphysics of
flame dynamics in the presence of a magnetic field.

In this paper, we present a linear-stability analysis of the
hydrodynamic flame instability, or Landau-Darrieus instability
\citep{landauflameprop,darrieusunpublished} in the presence of a
magnetic field.

\subsection{Landau-Darrieus Instability}

\begin{figure}
\plotone{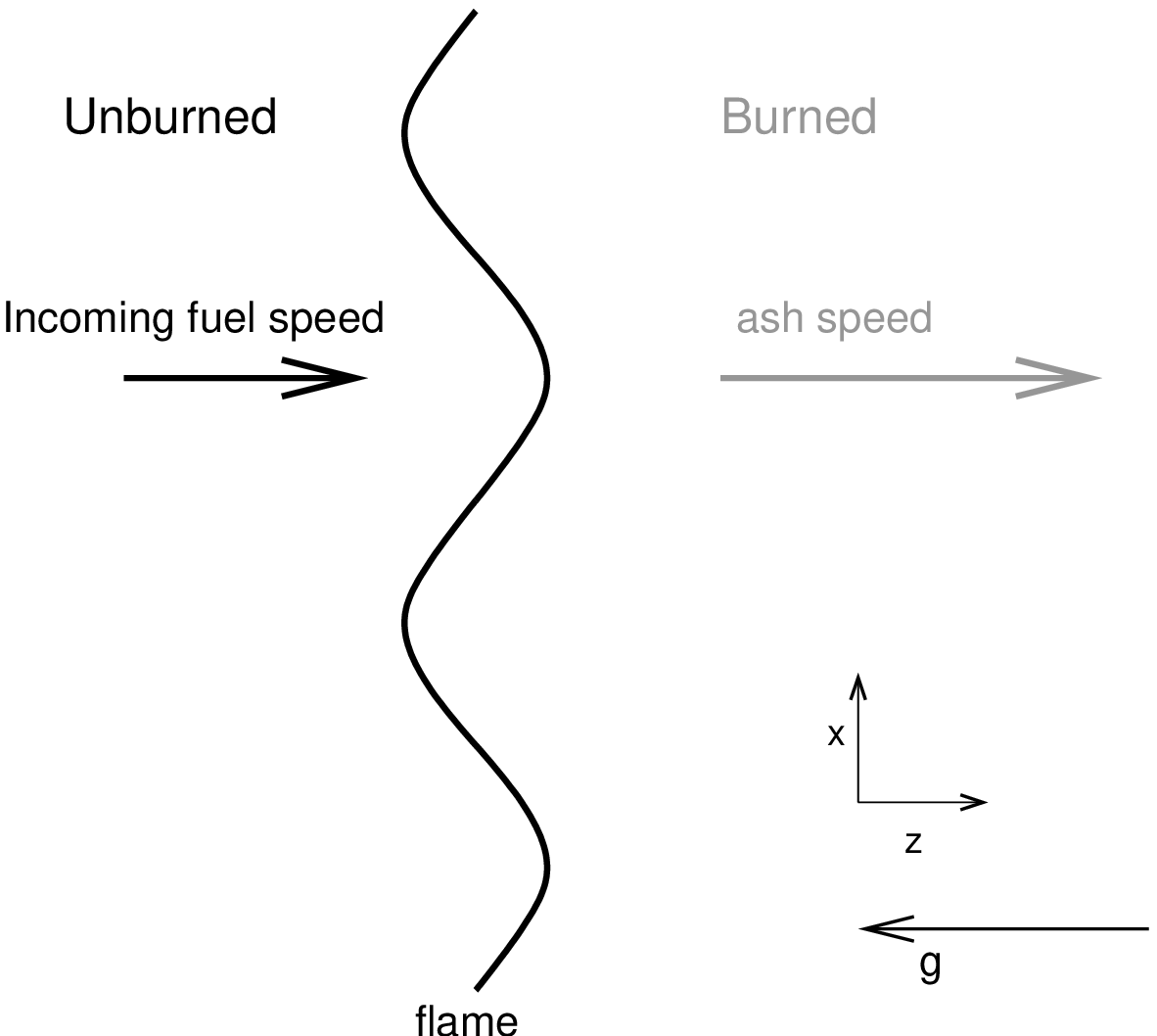}
\caption{A sketch of the physical setup of the Landau-Darrieus
instability in the frame of the flame front. }
\label{fig:ldsketch}
\end{figure}

The physical scenario of the Landau-Darrieus instability is sketched
in Fig.\ref{fig:ldsketch}, where burning proceeds in a subsonic wave (a
deflagration wave, or a flame), propagating as a plane.  For simplicity,
we consider the frame of the unperturbed flame.  The flame structure
itself is not modelled here; instead, we consider the flame to be
an infinitesimally thin interface propagating at a known speed,
$S_l$, into the fuel.   In the original Landau-Darrieus derivation,
the flame speed was assumed to be so much slower than the sound speed
that the unburned fuel and burned ash regions could be assumed to have
incompressible flows; in our astrophysical case, this assumption also
holds \citep{timmeswoosley}.   We use subscript $b$ to refer to quantities
in the burned region, and $u$ to refer to quantities in the unburned
region.  Fuel approaches the flame with a velocity equal to $S_l$.

\begin{figure}
\plotone{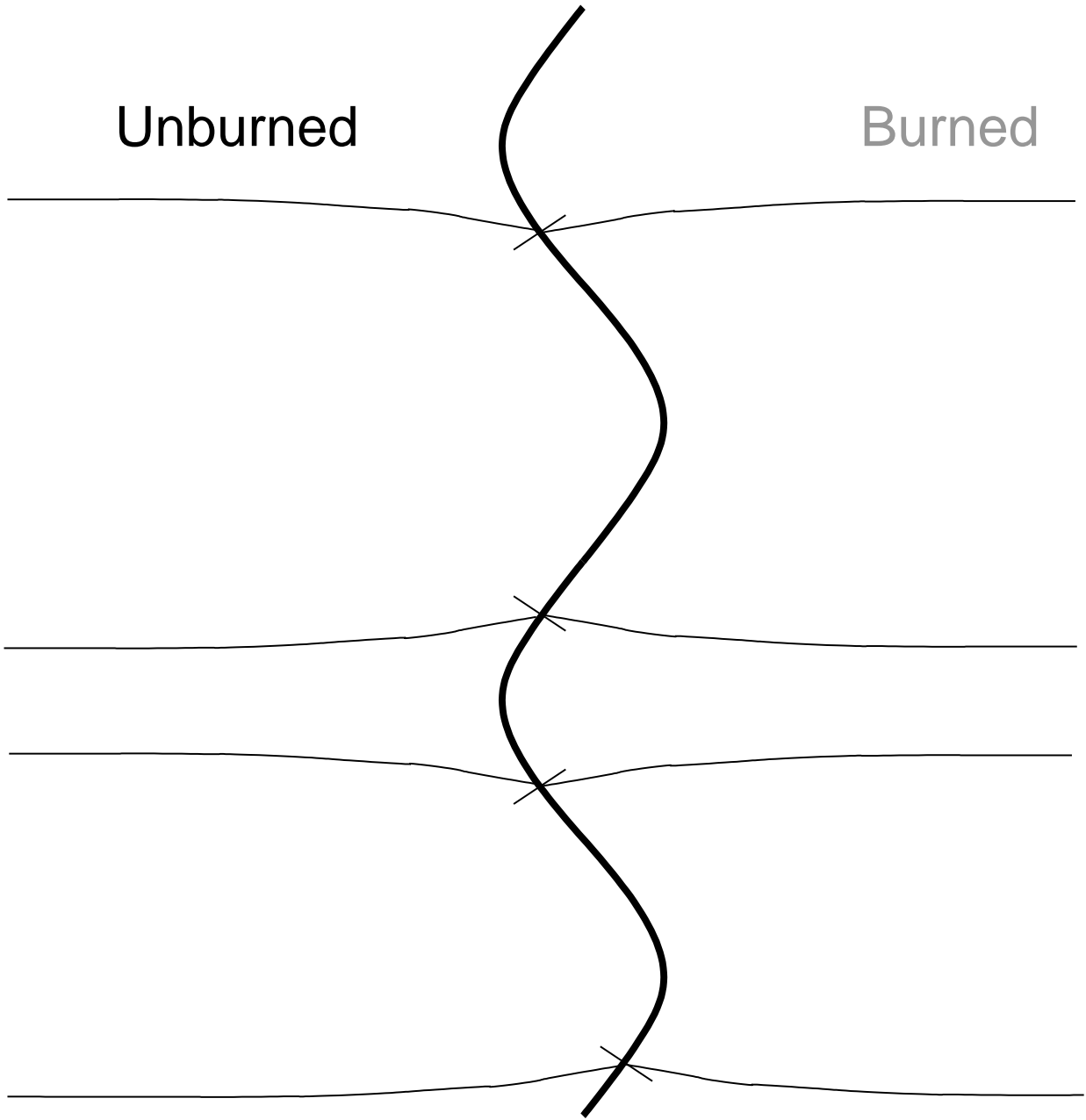}
\caption{Because the tangential component of velocity is constant across
the flame in the hydrodynamical case, while the normal component jumps, streamlines
crossing the flame are bent towards the flame normal.  Since  at large
distances the streamlines must remain parallel, the stream lines `fan out'
near flame surfaces curved towards the unburned gas; this locally reduces
the flow velocity, allowing the flame in that region to propagate further
ahead.  (Figure taken after  \cite{williams} Fig.~9.8).
}
\label{fig:ldmechanism}
\end{figure}

The reaction that drives the propagating flame is assumed to be exothermic
enough that the density behind the flame, $\rho_b$ is lower than the
density ahead of the flame, $\rho_u$.  This is expressed as a density
ratio, $\alpha = \rho_u/\rho_b > 1$.  In this case, streamlines are
refracted by the density jump across the flames, so that any initial
wrinkling of the flame drives further wrinkling (see for instance
\citealt{williams}, and Fig.~\ref{fig:ldmechanism}.) This instability,
driven by the density jump and mediated by hydrodynamic waves, is the
well-known Landau-Darrieus instability, and in the linear regime it has
an exponential growth rate of (\eg, \citealt{bychkovlibermanreview})
\begin{equation}
\frac{n}{k S_l} = \frac{\alpha}{\alpha + 1} \left ( \sqrt{ -\frac{g}{k S_l^2} \frac{\alpha^2-1}{\alpha^2} + \frac{\alpha^2 + \alpha - 1}{\alpha}} - 1 \right ),
\end{equation}
where $n$ is the growth rate, $k$ is the wavenumber of the perturbation,
and $g$ is the gravitational acceleration.   This instability is
thought to play roles in terrestrial combustion (see for instance
the references in \citealt{bychkovlibermanreview}), astrophysical
combustion \citep{hillebrandtniemeyer00}, and inertial confinement
fusion \citep{icfablation}, as well as at QCD and electroweak transition fronts
\citep{qcdwalls,burningbubbles} with possible astrophysical application
to strange stars \citep{strangepulsar}, polymerization fronts \citep{polymerizationfronts}
and evaporation fronts such as of liquid metals \citep{metalboiling}.

The wrinkling perturbs the pressure and velocities on either side of the
flame.  If the flame's speed into the fuel is taken to be fixed at
its planar value (the applicability of this assumption to astrophysical
flames is considered in \citealt{flamecurvature}), the velocity at which
the flame wrinkles is equal to the perturbed fluid velocities at the
interface -- \eg, the flame is advected by the perturbed velocity.

This hydrodynamical Landau-Darrieus instability has been
analytically examined in further detail in the astrophysical context
\citep{COld,blinnsaswoos,ldfractal}.  In this work, we extend the
derivation of the classical Landau-Darrieus instability to the case
of the presence of a magnetic field.  In \S\ref{sec:bparallel}, we
consider a background magnetic field oriented along the direction
of flame propagation.   In \S\ref{sec:bperpendicular}, we consider
a background magnetic field oriented transverse to the direction of
propagation.  In \S\ref{sec:localspeed} we examine the self-consistency
of assuming a fixed flame propagation speed when thermal diffusion will
preferentially operate along magnetic field lines, In \S\ref{sec:markstein}
we examine the effect of non-constancy of flame speed due to curvature,
and we conclude in \S\ref{sec:conclusions}.

% ============================================================================
%  field parallel to flame propagation
% ============================================================================
\section{FIELD ALONG DIRECTION OF FLAME PROPAGATION}
\label{sec:bparallel}
\subsection{Perturbation equations}
\label{sec:bparallelpert}

Here we derive the linear growth rate of the magnetic Landau-Darrieus
instability.  We follow closely the approach and notation of
\cite{chandra}.  We begin with the equations of incompressible
magnetohydrodynamics in an inviscid, perfectly conducting fluid:
\begin{eqnarray}
\label{eq:momentum}
\frac{\partial U_i}{\partial t} + U_j \frac{\partial U_i}{\partial x_j} 
            - \frac{\magmu \magH_j}{4 \pi \rho} \left ( \frac{\partial \magH_i}{\partial x_j} - \frac{\partial \magH_j}{\partial x_i}\right) & = &
            -\frac{1}{\rho} \frac{\partial p}{\partial x_i}  + g_i \frac{\delta \rho}{\rho}\\
\label{eq:induction}
\frac{\partial \magH_i}{\partial t} + \frac{\partial}{\partial x_j} \left ( U_j \magH_i - \magH_j U_i \right ) & = & 0 \\
\label{eq:incompressible}
\frac{\partial U_i}{\partial x_i} & = & 0 \\
\label{eq:nomonopoles}
\frac{\partial \magH_i}{\partial x_i} & = & 0
\end{eqnarray}
where $x_i$ is the $i$-coordinate, $U_i$ is the velocity in the $i$-coordinate
direction, $\rho$ is density, $\delta\rho$ is any fluctuation in the density,
$p$ is pressure, $\magH$ is the magnetic field, $g$ is the gravitational
acceleration, and summation over repeated indicies is implied.
We assume constant density states
($\delta\rho = 0$) on the burned and unburned sides, with both gravity
and a magnetic field pointing in the direction of propagation (which we
take here to be ${\bf{\vec{z}}}$).
We consider velocity, magnetic fields, and pressure of the form
\begin{eqnarray}
\label{eq:velperts}
{\bf{U}} & = & (u, v, w + W) \quad u,v,w \ll W, \\
\label{eq:magperts}
{\bf{\magH}} & = & (\magh_x, \magh_y, \magh_z + \magH) 
       \quad \magh_x,\magh_y,\magh_z \ll \magH, \\
       p & = & P + \delta p \quad \delta p \ll P,
\end{eqnarray}
where $P$, $\magH$, and $W$ are constant within each region.  The velocity
$W_u$ with which the fuel approaches the flame is taken to be $S_l$,
the flame speed, so that the unperturbed flame would be motionless.
We will assume that all velocities are highly non-relativistic, so that
displacement currents and relativistic effects may be neglected.  We also
assume that the flame and flow velocities are much less than the sound
speed, and thus may consider incompressible flow.  These assumptions
are all appropriate for the slow flames in highly degenerate material
that we consider here.

We then expand the MHD equations to linear order in the perturbed quantities, writing out
explicitly the components of the vector equations.   The linearization is described
in more detail in Appendix~\ref{sec:app:linearization}.   We obtain:
\begin{eqnarray}
\label{eq:xmomentumpert}
\frac{\partial u}{\partial t} + W \frac{\partial u}{\partial z} 
    - \frac{\magmu \magH}{4 \pi \rho} \left ( \frac{\partial \magh_x}{\partial z} - \frac{\partial \magh_z}{\partial x} \right )& = & 
    - \frac{1}{\rho}\frac{\partial}{\partial x} \delta p \\
\label{eq:ymomentumpert}
\frac{\partial v}{\partial t} + W \frac{\partial v}{\partial z} 
    - \frac{\magmu \magH}{4 \pi \rho} \left ( \frac{\partial \magh_y}{\partial z} - \frac{\partial \magh_z}{\partial y} \right )& = & 
    - \frac{1}{\rho}\frac{\partial}{\partial y} \delta p \\
\label{eq:zmomentumpert}
\frac{\partial w}{\partial t} + W \frac{\partial w}{\partial z} & = & 
    - \frac{1}{\rho}\frac{\partial}{\partial z} \delta p,
\end{eqnarray}
\begin{eqnarray}
\label{eq:xinductionpert}
\frac{\partial \magh_x}{\partial t} + W \frac{\partial \magh_x}{\partial z} & = & \magH \frac{\partial u}{\partial z} \\
\label{eq:yinductionpert}
\frac{\partial \magh_y}{\partial t} + W \frac{\partial \magh_y}{\partial z} & = & \magH \frac{\partial v}{\partial z} \\
\label{eq:zinductionpert}
\frac{\partial \magh_z}{\partial t} + W \frac{\partial \magh_z}{\partial z} & = & \magH \frac{\partial w}{\partial z} ,
\end{eqnarray}
\begin{eqnarray}
\label{eq:incompressiblepert}
\frac{\partial u}{\partial x} + \frac{\partial v}{\partial y} + \frac{\partial w}{\partial z} & = & 0\\
\label{eq:nomonopolespert}
\frac{\partial \magh_x}{\partial x} + \frac{\partial \magh_y}{\partial y} + \frac{\partial \magh_z}{\partial z} & = & 0.
\end{eqnarray}

We now consider normal modes, where all perturbed quantities have $x$, $y$, and $t$ dependencies of the form
$\exp(i k_x x + i k_y y + n t)$.   In that case, and using $D$ to denote $d/dz$, the equations become
\begin{eqnarray}
\label{eq:xmomentummode}
\nwd u - \frac{\magmu \magH}{4 \pi \rho} \left ( D \magh_x - i k_x \magh_z \right ) & = & -\frac{i k_x}{\rho} \delta p \\
\label{eq:ymomentummode}
\nwd v - \frac{\magmu \magH}{4 \pi \rho} \left ( D \magh_y - i k_y \magh_z \right ) & = & -\frac{i k_y}{\rho} \delta p \\
\label{eq:zmomentummode}
\nwd w                                                               & = & -\frac{1}{\rho} D \delta p \\
\label{eq:xinductionmode}
\nwd \magh_x & = & \magH D u \\
\label{eq:yinductionmode}
\nwd \magh_y & = & \magH D v \\
\label{eq:zinductionmode}
\nwd \magh_z & = & \magH D w \\
\label{eq:incompressiblemode}
D w & = & - \left ( i k_x u + i k_y v \right ) \\
\label{eq:nomonopolesmode}
D \magh_z & = & - \left ( i k_x \magh_x + i k_y \magh_y \right ).
\end{eqnarray}

Adding $-i k_x$ times Eq.~\ref{eq:xmomentummode} to $-i k_y$ times Eq.~\ref{eq:ymomentummode}
and using Eqs.~\ref{eq:incompressiblemode} and \ref{eq:nomonopolesmode} gives us
\begin{equation}
\nwd D w - \frac{\magmu \magH}{4 \pi \rho} ( D^2 - k^2 ) \magh_z  =  -\frac{k^2}{\rho} \delta p.
\end{equation}
Applying $D$ again and substituting $\delta p$ from Eq.~\ref{eq:zmomentummode} gives 
\begin{equation}
\nwd (D^2 - k^2) w - \frac{\magmu \magH}{4 \pi \rho} ( D^2 - k^2 ) D \magh_z  =  0.
\end{equation}
Applying $\nwd$ to this and applying a component of the induction equation, Eq.~\ref{eq:zinductionmode}
gives
\begin{equation}
\left (\nwd^2 - (a D)^2\right) (D^2 - k^2) w =  0
\end{equation}
where $a^2 \equiv \magmu \magH^2/(4 \pi \rho)$ is the square of the Alfv\'en speed.   Similar equations for the other variables are:
\begin{eqnarray}
\left (\nwd^2 - (a D)^2\right) (D^2 - k^2) D \delta p & = &  0 \\
\left (\nwd^2 - (a D)^2\right) (D^2 - k^2) \nwd \magh_z & = &  0.
\end{eqnarray}
We note that we are free to consider rotations around the $\bf{\vec{z}}$
axis such that $k_y$ is zero and $k = k_x$.   If we do so, then $\magh_y$ and $v$
decouple from the other quantities and we need not consider them 
further.  Then $u = (i/k) Dw$ and $\magh_x = (i/k) D \magh_z$, completing our
set of equations.

The $z$-dependence of our perturbed variables, then, must be of the form
\begin{equation}
\coefa e^{k z} + \coefb e^{-k z} + \coefc e^{-\frac{n}{W}z} + \coefd e^{-\frac{n}{W+a} z} + \coefe e^{-\frac{n}{W-a} z}.
\end{equation}
where the $C^{(i)}$ are constants within each of the burned and
unburned regions, and represent the
amplitudes of waves which travel along the $\bf{\hat{z}}$-direction
with speeds $-n/k$, $n/k$, $W$, $W+a$, and $W-a$, respectively.   
Since the perturbed quantities are small (order $\epsilon$, as per
Appendix~\ref{sec:app:linearization}), these $C^{(i)}$ must also be small
(order $\epsilon$) quantities.

In the case of flames propagating at the Alfv\'en speed $W = a$ in either
the unburned or burned medium, the terms $(\nwd^2 - (a D)^2)$ become
$n (n + (a + W) D)$, and thus only a wave corresponding to 
$\coefd$, and not $\coefe$, exists in that region.

The unburned state will contain only waves that propagate from the flame
to $-\infty$; \ie, those with negative speed.  Since $W$
and $a$ are positive, and we are only interested in modes with the
real component of $n$ positive, this leaves the first wave, and
the fifth if the flame is slower than the Alfv\'en speed: $W_u < a_u$.
The perturbations in the unburned state are then:
\begin{eqnarray}
\delta p_u  & = & -\frac{\rho_u (n + W_u k)^2}{k^2} \coefa_u e^{k z} - \rho_u a_u^2 \coefe_u e^{-\frac{n}{W_u - a_u} z} \\
{\magh_z}_u & = & \magH \coefa_u e^{k z} + \magH \coefe_u e^{-\frac{n}{W_u - a_u} z} \\
w_u   & = & \frac{(n + W_u k)}{k} \coefa_u e^{k z} + a_u  \coefe_u e^{-\frac{n}{W_u - a_u} z} \\
{\magh_x}_u   & = & i \magH \coefa_u e^{k z} - i \magH \frac{n}{k(W_u-a_u)} \coefe_u e^{-\frac{n}{W_u - a_u} z} \\
u_u   & = & i \frac{(n + W_u k)}{k} \coefa_u e^{k z} - i \frac{a_u}{k} \frac{n}{W_u-a_u} \coefe_u e^{-\frac{n}{W_u - a_u} z} 
\end{eqnarray}
where these expressions were taken by choosing the parameters for ${\magh_z}_u$, calculating $w_u$
from the induction equation and then $\delta p$ from the momentum equation, and then finding
${\magh_x}_u$ from ${\magh_z}_u$ using the divergence condition on the magnetic field and similarly
using incompressibility to find $u_u$ from $w_u$. 

The same process can be repeated for the burned state.  This gives us:
\begin{eqnarray}
\delta p_b & = & -\frac{\rho_b (n - W_b k)^2}{k^2} \coefb_b e^{-k z} - \rho_b a_b^2 \coefd_b e^{-\frac{n}{W_b + a_b} z} 
 - \rho_b a_b^2 \coefe_b e^{-\frac{n}{W_b-a_b} z} 
\\
{\magh_z}_b & = & \magH \coefb_b e^{-k z} + \magH \coefd_b e^{-\frac{n}{W_b + a_b} z} 
+ \magH \coefe_b e^{-\frac{n}{W_b-a_b} z} 
\\
w_b & = & - \frac{(n - W_b k)}{k} \coefb_b e^{-k z} - a_b \coefd_b e^{-\frac{n}{W_b + a_b} z} 
+ a_b \coefe_b e^{-\frac{n}{W_b-a_b} z} 
\\
{\magh_x}_b & = & - i \magH \coefb_b e^{-k z} - i \magH \frac{n}{k(W_b + a_b)} \coefd_b e^{-\frac{n}{W_b + a_b} z} 
 - i \magH \frac{n}{k(W_b - a_b)} \coefe_b e^{-\frac{n}{W_b-a_b} z} 
\\
u_b & = & i \frac{(n - W_b k)}{k} \coefb_b e^{-k z} + i \frac{a_b}{k} \frac{n}{W_b + a_b} \coefd_b e^{-\frac{n}{W_b + a_b} z} 
 - i \frac{a_b}{k} \frac{n}{W_b - a_b} \coefe_b e^{-\frac{n}{W_b-a_b} z} 
.
\end{eqnarray}
Note the absence of $\coefc e^{-(n/W) z}$ terms; for consistency between
the momentum equation and the magnetic field divergence condition, $\coefc$
must be zero.

\subsection{Jump Conditions}
\label{sec:bparallelbcs}

Having derived the form of the perturbations in each region, we
must provide the matching conditions at the flame, treated here as a
discontinuity, to solve for the perturbations and their growth rate.
Jump conditions in an inviscid, perfectly conducting magnetic fluid  
are (see for instance \cite{shu2,tidman}):
\begin{eqnarray}
\left [ \magH_n \right ] & = & 0 \\
\left [ \rho U_n \right ] & = & 0 \\
\rho U_n \left [ {\bf U_t} \right ] - \frac{\magmu \magH_n}{4 \pi} \left [ \bf{\magH_t} \right ] & = & 0 \\
\rho U_n \left [ U_n \right ] + \left [ p + \frac{\magmu \magH_t^2}{8 \pi} \right ] & = & g z_f \left [ \rho \right ]\\
\left [ U_n {\bf \magH_t} \right ] - \magH_n \left [ {\bf{U_t}} \right ] & = & 0,
\end{eqnarray}
where $[f]$ represents the jump in quantity $f$ across the discontinuity,
a subscript $n$ refers to the component normal to the discontinuity,
and subscript $t$ refers to components tangential to the discontinuity.

The velocity normal to the discontinuity, $U_n$, must be the flame speed
$S_l$, which is prescribed; here we are considering the fixed flame
speed $S_l = W_u$.  Since the interface is moving with respect to the
far upstream fluid at $W + w_i$, where $w_i$ is the $w$-velocity at the
interface, the flame must be being advected --- that is, wrinkled ---
with a velocity of $W + w_i - U_n = w_i$.

The directions tangential to the flame can be determined by taking
the derivatives of the flame position $z_f$, \eg $\partial z_f /\partial x$
and $\partial z_f/\partial y$.   Because the flame position
(in the unperturbed flame frame) is being wrinkled by the perturbed
fluid velocity, and is itself a perturbed quantity $z_f \sim e^{i k_x x +
i k_y y + n t}$, we can express $z_f$ in terms of $w_i$:
\begin{eqnarray}
\frac{\partial}{\partial t} z_f & = & w_i \\
z_f & = & \frac{w_i}{n}.   
\end{eqnarray}
where $w_i = w(z_f)$.  
The difference between $w(z_f)$ and $w(0)$
is of second order of smallness, so we consider $w_i = w(0)$.
This gives us the normal and two tangential
directions, to linear order:
\begin{eqnarray}
{\bf{\hat{n}}}   & = & \left ( -\frac{i k_x}{n} w_i, -\frac{i k_y}{n} w_i, 1 \right ) \\
{\bf{\hat{t}_1}} & = & \left ( 1, 0, \frac{i k_x}{n} w_i \right ) \\
{\bf{\hat{t}_2}} & = & \left ( 0, 1, \frac{i k_y}{n} w_i \right ).
\end{eqnarray}
We note again that we can ignore the $y$-component, and $\bf{\hat{t}_2}$.
Then we have
\begin{eqnarray}
\magH_n & = & \magh_z + \magH \\
\magH_t & = & \magh_x + \frac{i k}{n} w_i \magH \\
U_t & = & u + \frac{i k}{n} w_i W \\
U_n & = & W.
\end{eqnarray}

To zeroth order in the perturbed quantities, then, the jump conditions
are:
\begin{eqnarray}
\left [ \magH \right ] & = & 0 \\
\left [ \rho W \right ] & = & 0 \\
\rho W \left [ W \right ] + \left [ P \right ] & = & 0 \\
\end{eqnarray}
and linear order gives us 
\begin{eqnarray}
\label{eq:hzcontin}
\left [ \magh_z \right ] & = & 0 \\
\label{eq:momtcontin}
\rho W \left [ u + \frac{i k}{n} w_i W \right ] - \frac{\magmu \magH}{4 \pi} \left [ \magh_x \right ] & = & 0 \\
\label{eq:momncontin}
\left [ \delta p \right ] & = & \frac{g w_i}{n} \left [ \rho \right ] \\
\label{eq:inductjump}
\left [ W \magh_x \right ] - \magH \left [ u \right ] & = & 0.
\end{eqnarray}
A further condition, 
\begin{equation}
\label{eq:noholes}
\left [ w_i \right ] = 0 ,
\end{equation}
simply states that each side of the discontinuity travels with the
same velocity.

The continuity of $\magh_z$
across the interface implies (through the induction equation and the
divergence-free property of the magnetic field) Eq.~(\ref{eq:inductjump}),
so that Eq.~(\ref{eq:inductjump}) it may be discarded in favor of
Eq.~(\ref{eq:hzcontin}).   This leaves us with four jump conditions:
Eqns.~(\ref{eq:hzcontin}), (\ref{eq:momtcontin}), (\ref{eq:momncontin}),
and (\ref{eq:noholes}).

We wish to solve for the perturbed quantities, so it is the linear-order
jump conditions that we are interested in.   The zeroth-order conditions
give us some restrictions on the quantities ($W$, $\rho W$,
and $\magH$) that appear in the linear-order conditions.

We apply these conditions at $z = 0$, as $z_f$ itself is a perturbed
quantity, and as with $w_i$, the difference between applying the boundary conditions at
$z = 0$ and $z = z_f$ will be of second order in smallness.  Thus, for any
jump condition $[f] = 0$, we set $f_b(0) - f_u(0) = 0$.  These boundary
conditions can then be expressed as a series of linear equations in the
coefficients of the $z$ dependences, ${\cal M} \cdot (\coefa_u, \coefb_b, \coefd_b,
\coefe_u, \coefe_b)^T = 0$.   The matrix ${\cal M}$ is
\begin{equation}
\tiny
\label{eq:parallelsystem}
\left [ \matrix{ -\left( 1 + \bar{n} \right)  & \left( \alpha - \bar{n} \right)  & -\sqrt{\alpha} \left( {\sqrt{\alpha}} + \bar{a}_u \right) & 
         1- \bar{a}_u & {\sqrt{\alpha}} \bar{a}_u \cr 
 -\left( 1 - \left( -2 + {\bar{a}_u}^2 \right)  \bar{n} + {\bar{n}}^2 \right) & \left( {\alpha}^2 - 2 \alpha \bar{n} + \bar{n} \left( {\bar{a}_u}^2 + \bar{n} \right)  \right)   & \frac{-\left( {\sqrt{\alpha}} + \bar{a}_u \right)  \left( {\alpha}^2 - {\bar{n}}^2 \right)}{ {\sqrt{\alpha}} } & \left( -1 + \bar{a}_u \right)  \left( -1 + {\bar{n}}^2 \right) & \frac{\bar{a}_u \left( {\alpha}^2 - {\bar{n}}^2 \right) }{\sqrt{\alpha} } \cr 
\left( 1 + \bar{n} \right)  \left( \bar{g} + \bar{n} + {\bar{n}}^2 \right) & 
-\left( \frac{\left( \alpha - \bar{n} \right)  \left( \bar{g} + \left( \alpha - \bar{n} \right)  \bar{n} \right) }{\alpha} \right) &
 -\left( \frac{\left( {\sqrt{\alpha}} + \bar{a}_u \right)  \left( -\bar{g} + {\sqrt{\alpha}} \bar{a}_u \bar{n} \right) }{\sqrt{\alpha}} \right)  &
 \left( -1 + \bar{a}_u \right)  \left( \bar{g} + \bar{a}_u \bar{n} \right)  & 
-\left( \frac{\bar{a}_u \left( \bar{g} + {\sqrt{\alpha}} \bar{a}_u \bar{n} \right)}{{\sqrt{\alpha}} } \right)  \cr 
- \bar{a}_u & \bar{a}_u & \left( {\sqrt{\alpha}} + \bar{a}_u \right)  & -\left( -1 + \bar{a}_u \right) &  \bar{a}_u \cr 
} \right ]
\end{equation}
where $\bar{n} = n (k S_l)^{-1}$ is the dimensionless growth rate,  $\bar{g}
= g (k S_l^2)^{-1}$ is the dimensionless acceleration due to gravity, and
$\bar{a}_u = a_u  S_l^{-1}$ is the dimensionless Alfv\'en speed in the fuel.
We have expressed all quantities in terms of the unburned state,
using $\rho_u = \alpha \rho_b$, $S_l = W_u$, $\alpha W_u = W_b$, and
$\sqrt{\alpha} a_u = a_b$.

\subsection{Growth Rates}

\begin{figure}
\plotone{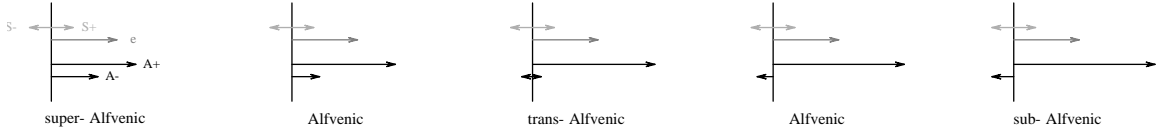}
\caption{The waves and their speeds propagating from the flame (vertical line), following
         \cite{anderson}, in the case of the field parallel to flame propagation.   The waves shown, top to bottom, are 
         $\coefa$/$\coefb$, $\coefc$, $\coefd$, and $\coefe$.   
         From left to right, the states are: super-Alfv\'enic ($\bar{a}_u < 1$);
         `switch on' Alfv\'enic, ($\bar{a}_u = 1$); trans-Alfv\'enic
         ($1 < \bar{a}_u < \sqrt{\alpha}$); `switch off' Alfv\'enic, 
         ($\bar{a}_u = \sqrt{\alpha}$); and sub-Alfv\'enic ($\bar{a}_u > \sqrt{\alpha}$).} 
\label{fig:waves}
\end{figure}

Depending on the relation of the flame and Alfv\'en velocities,
different waves will be able to propagate away from the flame;
this will determine the stability properties of the flame.   
The different regimes are shown in Fig.~\ref{fig:waves}.

\subsubsection{Super-Alfv\'enic flames: $W_u > a_u$; $W_b > a_b$}

In this case, no Alfv\'en waves can travel upstream, so that $\coefe_u$
must be zero.   Then we have four constraints on four waves, and the
system of equations represented by Eq.~(\ref{eq:parallelsystem}) has a
solution if the determinant of
$\cal M$ is zero;
\begin{eqnarray}
\left( {\sqrt{\alpha}} + \bar{a}_u \right)  \left( -{\alpha}^2 - {\bar{n}}^2 + \alpha \left( {\bar{a}_u}^2 + 2 \bar{n} \right)  \right)  \times \nonumber \\
  \left( {\alpha}^2 \left( 1 + \bar{n} \right)  + \left( 1 + \bar{n} \right)  \left( \bar{g} - {\bar{n}}^2 \right)  - 
    \alpha \left( 1 + \bar{g} + 3 \bar{n} - 2 {\bar{a}_u}^2 \bar{n} + \bar{g} \bar{n} + 
       3 {\bar{n}}^2 + {\bar{n}}^3 \right)  \right) & = & 0
\end{eqnarray}
Further simplification, and the removal of two trivial solutions --- $\bar{n} = \alpha \pm \sqrt{\alpha} \bar{a}_u$
--- that lead to no actual wrinkling of the interface leaves us with:
\begin{equation}
\bar{n}^3 + \frac{1 + 3 \alpha}{1 + \alpha} \bar{n}^2 
   - \frac{\alpha^2 + \alpha (2 \bar{a}_u^2 - \bar{g} - 3) + \bar{g}}{1 + \alpha} \bar{n}
   - \frac{\alpha^2 - \alpha(\bar{g}+1) + \bar{g}}{1 + \alpha} = 0.
\label{eq:bparsupergrowth}
\end{equation}
In the limit of $\bar{a}_u \rightarrow 0$, this reproduces the well-known
Landau-Darrieus result.

Absent gravity, we see that the constant term in the cubic is necessarily
negative, so that a positive real solution always exists for $\bar{n}$
--- that is, the interface is always unstable.   The magnetic field
cannot itself stabilize the flame in this case.  One may examine the
case of marginal stability in the presence of gravity by considering
$\bar{n} = 0$.  This requires $\alpha^2 - \alpha(\bar{g}+1) + \bar{g}
= 0$.  Since this does not include terms in $\bar{a}_u$, one must recover
the Landau-Darrieus result; namely, that for stabilizing acceleration
the condition of marginal stability is $\alpha = \bar{g}$, with larger
$\alpha$ being unstable.

\begin{figure}
\plotone{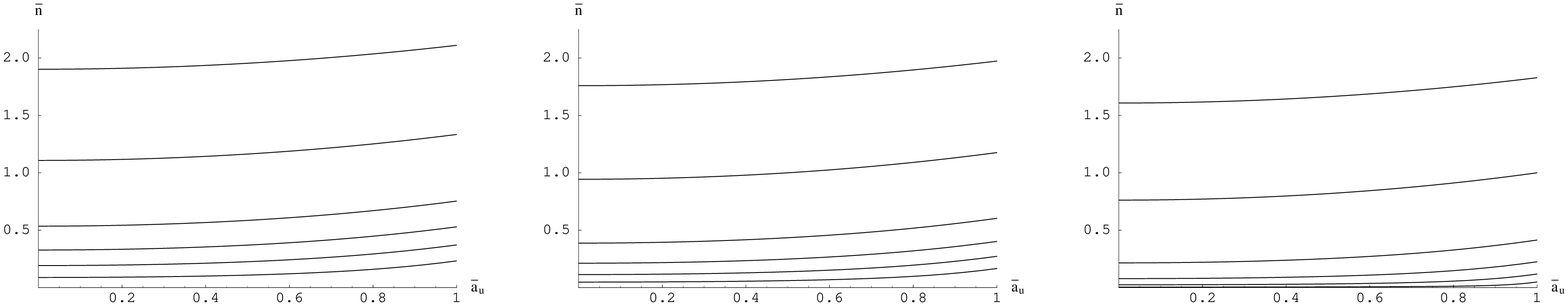}
\caption{The growth rate for a parallel magnetic field when the flame
is everywhere super-Alfv\'enic.   We plot the maximum real
part of the scaled growth rate
($\bar{n}$) as a function of $\bar{a}_u$, with $\bar{g} = -1$ (destabilizing)
on the left, $\bar{g} = 0$ in the center, and $\bar{g} = +1$ (stabilizing)
on the right.  Lines are plotted for, top to bottom, $\alpha$ = 8, 4, 2, 1.5, 1.25, 1.1.
}
\label{fig:parallel-super}
\end{figure}

The maximum growth rate is plotted in this case for varying $\bar{g}$,
$\alpha$, and $\bar{a}_u$ in Fig.~\ref{fig:parallel-super}.   We see that
in this case, the instability is slightly enhanced over the non-magnetic
($\bar{a}_u = 0$) case.

\subsubsection{`Switch On' Alfv\'enic flames: $W_u = a_u$; $W_b > a_b$}

When the upstream flow is exactly Alfv\'enic, a tangential field
may ``switch on'' from zero in the upstream state to non-zero in
the downstream state;  such discontinuities are called ``switch on''
discontinuities (\eg, \citealt{anderson}).  If $W_u = a_u$, no $\coefe_u$
wave exists, as in the strictly super-Alfv\'enic case, and all else
remains the same; thus, the growth rate is the $\bar{a}_u \rightarrow 1$
limit of the super-Alfv\'enic case.

\subsubsection{Sub-Alfv\'enic flames: $W_u < a_u$; $W_b < a_b$}

In this case, upstream Alfv\'en waves may propagate, and thus
$\coefe_u$ may be non-zero but $\coefe_b$ must be zero.
Again, $\cal M$ becomes a square matrix and we may find the
solution of the linear equations by calculating the determinant:
\begin{eqnarray}
\left( -1 + \bar{a}_u - \bar{n} \right)  \left( \alpha + {\sqrt{\alpha}} \bar{a}_u - \bar{n} \right)  
    \left( {\alpha}^{\frac{5}{2}} + {\alpha}^2 \bar{n} - \bar{g} \bar{n} + {\bar{n}}^3 - 
      {\sqrt{\alpha}} \left( 1 + 2 \bar{a}_u \right)  \left( \bar{g} - {\bar{n}}^2 \right)  - \right . \nonumber \\
\left .      {\alpha}^{\frac{3}{2}} \left( 1 + \bar{g} + 2 \left( 2 + \bar{a}_u \right)  \bar{n} + {\bar{n}}^2 \right)  + 
      \alpha \left( \bar{g} \left( 2 + 2 \bar{a}_u + \bar{n} \right)  + 
         \bar{n} \left( 1 + 2 {\bar{a}_u}^2 + {\bar{n}}^2 + 2 \bar{a}_u \left( 1 + \bar{n} \right)  \right) 
         \right)  \right)  & = & 0.
\end{eqnarray}
Factoring out the two solutions ($\bar{n} = \bar{a}_u - 1$, $\bar{n} =
\sqrt{\alpha} \bar{a}_u + \alpha$) that result in trivial solutions,
one again has a cubic in $\bar{n}$,
\begin{eqnarray}
\label{eq:cubiclargeparallel}
\bar{n}^3 + 
\sqrt{\alpha}\frac{1 - \alpha + 2 \bar{a}_u + 2 \sqrt{\alpha} \bar{a}_u}{1 + \alpha} \bar{n}^2  && \nonumber  \\
+ \frac{\alpha^2 - 2 \alpha^\frac{3}{2} (2 + \bar{a}_u) - \bar{g} + \alpha(1 + 2 \bar{a}_u + 2 \bar{a}_u^2 + \bar{g})}{1 + \alpha} \bar{n} 
\nonumber && \\
+ \frac{(\sqrt{\alpha}-1)\sqrt{\alpha}\left ( \alpha  + \alpha^\frac{3}{2} + \bar{g}(1 + 2 \bar{a}_u - \sqrt{\alpha})\right)}{1 + \alpha}  & = & 0.
\end{eqnarray}
Note that in this case, even absent a gravitational acceleration $g$,
the constant term in the cubic is positive, meaning that no 
positive real solution need exist.

If we express this in terms of $n$ rather than $\bar{n}$, and take
the limit $W_u \rightarrow 0$ (\eg, the interface does not propagate
into the fluid),  one can reproduce the results of the Rayleigh-Taylor
instability in the presence of a vertical magnetic field, as discussed
in \S~96 of \cite{chandra}.

\begin{figure}
\plotone{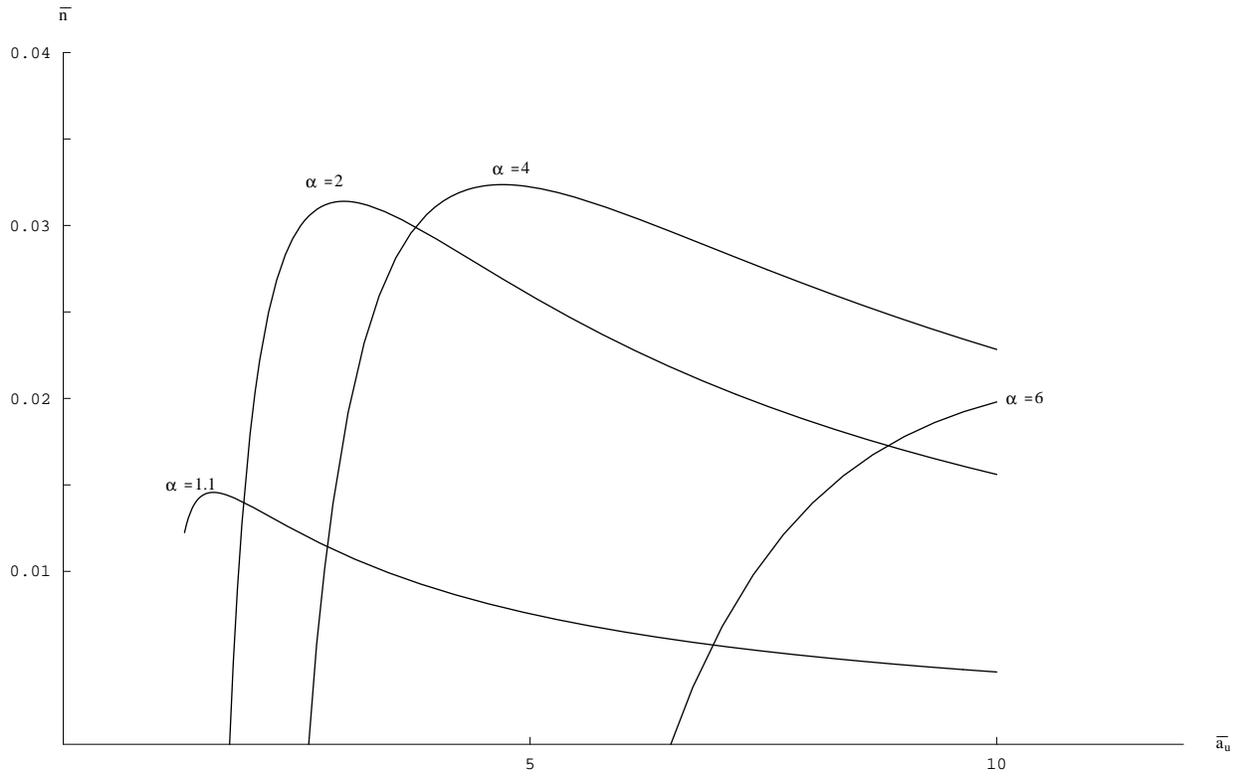}
\caption{The growth rate for a parallel magnetic field in the
sub-Alfv\'enic case with destabilizing gravity; with zero or stabilizing
gravity, the flame is stable.   We plot the maximum real part of the scaled
growth rate ($\bar{n}$) as a function of $\bar{a}_u$, with $\bar{g} = -1$.
Lines are plotted for, left to right,
$\alpha$ = 1.1, 2, 4, 6.  
}
\label{fig:parallel-sub}
\end{figure}

\begin{figure}
\plotone{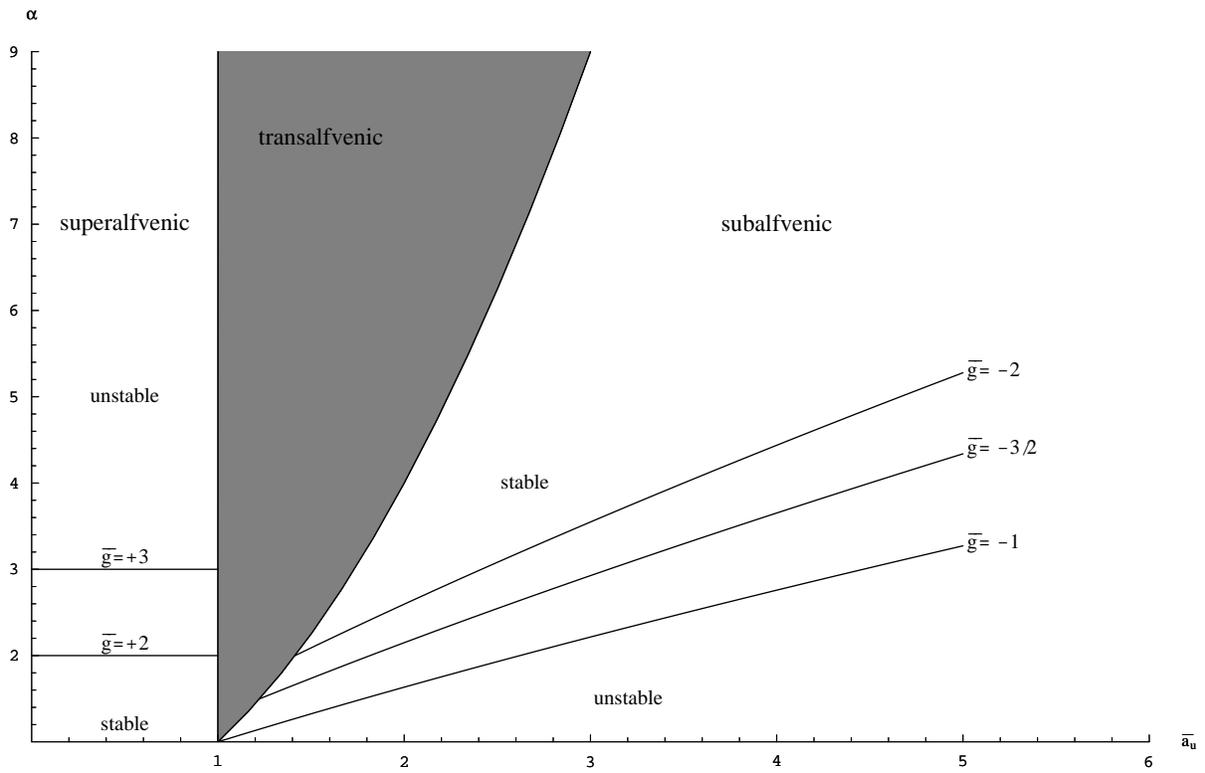}
\caption{Stability boundaries in $\alpha$ for the case of a parallel
magnetic field as a function of $\bar{a}_u$.   Note that for the
superalfv\'enic case, the region above the curves is unstable, and for
the subalfv\'enic case, the region below the curves is unstable.
}
\label{fig:parallel-stability}
\end{figure}

The scaled growth rate $\bar{n}$ as a function of $\alpha$ is shown
in Fig.~\ref{fig:parallel-sub}.   The instability can be 
completely suppressed by the magnetic field.    The region of stability 
is plotted in Fig.~\ref{fig:parallel-stability}.

\subsubsection{`Switch Off' flames: $a_u > W_u$, $a_b = W_b$}

When the downstream flow is Alfv\'enic, a tangential field may ``switch
off'' in the downstream case; such discontinuities are called ``switch off''
discontinuities.   

If $W_b = a_b$, there can be no $\coefe_b$ wave, and all proceeds as
in the sub-Alfv\'enic case with $\bar{a}_u \rightarrow \sqrt{\alpha}$.

\subsubsection{Discussion}

Given that a sufficiently strong magnetic field suppresses the
instability, it is somewhat surprising that while the flame is still
super-Alfv\'enic, the growth rate of the instability actually increases
with increasing magnetic field strength.

\begin{figure}
\plotone{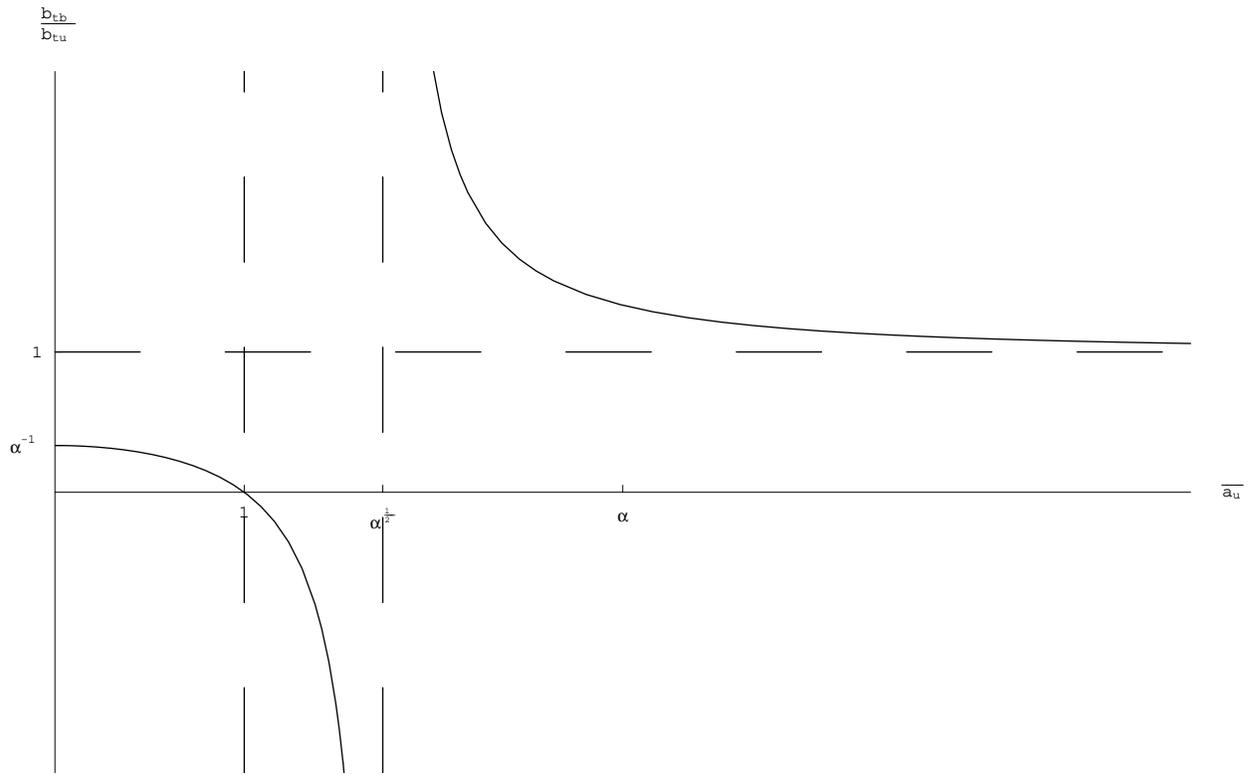}
\caption{Plotted is the ratio of the tangential magnetic field at the
interface in the burned fluid to that in the fuel, ${B_t}_b/{B_t}_u =
([B_t]/{B_t}_u) + 1$.   At $\bar{a}_u = 1$, the flame is a `switch off'
discontinuity; at $\bar{a}_u = \sqrt{\alpha}$, the flame is a `switch on'
discontinuity.
}
\label{fig:switchonoff}
\end{figure}

The reason for this is that the presence of a magnetic field allows
for a non-zero jump in the tangential velocity across the flame,
proportional to the jump in the tangential magnetic field component.
From the jump conditions, we can calculate the ratio of the tangential
magnetic fields across the interface as a function of $\bar{a}_u$;
this is shown in Fig.~\ref{fig:switchonoff}.   At the Alfv\'enic points,
the flame becomes a `switch on' or `switch off' discontinuity.  For the
super-Alfv\'enic flame ($\bar{a}_u < 1$), we see that the tangential
magnetic field decreases across the flame, and thus so does the
tangential velocity.   In particular, in this case one has
\begin{equation}
\frac{{U_t}_b}{{U_t}_u} = 1 - \frac{\alpha - 1}{\alpha - {\bar{a}_u}^2} {\bar{a}_u}^2 \frac{\bar{n}+2}{(\bar{n} + 1)^2}
\end{equation}
which is less than one.  Since the magnitude of the tangential velocity
decreases across the flame, the streamlines crossing the flame are bent
still closer to the normal than in the purely hydrodynamical case shown
in Fig.~\ref{fig:ldmechanism}, and the instability is modestly enhanced.

\subsubsection{Trans-Alfv\'enic flames: $W_u < a_u$; $W_b > a_b$}

In this case the coefficients $\coefd_b$, $\coefe_u$, and $\coefe_b$
may all be non-zero; we have a total of three Alfv\'en waves that may
propagate to infinity from the flame front.   Trans-Alfv\'enic flames
then have more possible waves leaving the flame surface than boundary
conditions to constrain them; there is no unique solution to the
initial-value problem.

In this case, the discontinuity is said to be non-evolutionary
(see for instance \citealt{anderson}, or a recent review by
\citealt{evolutionaryreview}.)   Analytically, these discontinuities have
no solution to the semi-steady time evolution of a small perturbation;
they must then undergo some finite-sized growth or change.   In the case
of an MHD shock, the consequences of nonevolutionarity are unclear;
numerical simulations of MHD shocks have been seen to spontaneously
break up into other discontinuities \citep{oscdisintegrate}, or stay
relatively steady over long periods \citep{ISMnonev}, depending on such
things as shock geometry and strength.   Numerical simulations will have
to be performed to examine the behavior of flames in these circumstances.

Simply considering the dissipative structure of an MHD shock structure
does not significantly reduce the range of parameters in which it
is non-evolutionary\citep{evolutionaryreview}.   It is possible that
adding the physics of the (compressible) flame structure to the jump
conditions as was done in \cite{pelceclavin82} or \cite{matalonmatkowsky}
for terrestrial non-magnetic flames may select out a particular solution
in this case.   This has never been done with a magnetic field, and is
well beyond the scope of this work.

It is possible, however, to constrain the possible evolution by
parameterizing the allowed solutions.   We consider taking the amplitude
of the wave $C^{(A-)}_u$ as a free parameter; the algebra simplifies if
one considers a scaled ratio of this wave to the other forward-going wave,
\begin{equation} \theta = (\alpha - 1)
\left [ (\bar{a}_u - 1) \frac{C^{(A-)}_u}{C^{(s-)}_u} + 1 \right ].
\end{equation}
Note that, depending on the phase difference between the waves, $\theta$
may be imaginary.

In this case, one obtains the following equation for the scaled growth rate,
\begin{equation}
\bar{n}^3 + \frac{4 \alpha - \theta}{1 + \alpha} \bar{n}^2 - 
\frac{\alpha^2 - \alpha(\bar{g} - 2 \bar{a}_u^2 + 3) + \bar{g}}{1 + \alpha} \bar{n}
+ \frac{\theta ( \bar{g}-\alpha)}{1 + \alpha} = 0,
\end{equation}
which reduces (as it must) to the superalfv\'enic relation if $C^{(A-)}_u = 0$.

\begin{figure}
\plotone{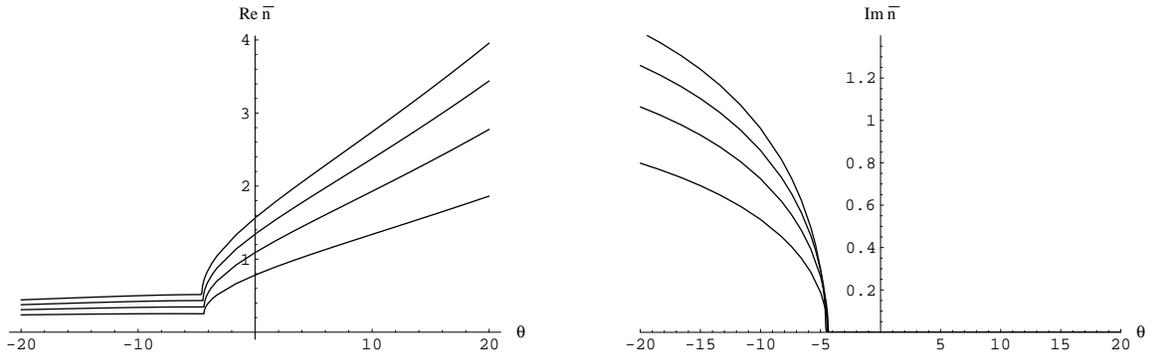}
\caption{On the left, the maximum real part of the scaled growth rate ($\bar{n}$) in the
trans-Alfv\'enic case, with $\bar{g} = 0$, $\bar{a}_u = 1.3$, and $\alpha$ (top to bottom) 5,4,3,2.
On the right, the corresponding imaginary component.   For any value of $\theta$ the flame is
unstable; the flame can only partially stabilize itself by sending forward fairly strong (several
times the amplitude of the $C^{(s-)}$ wave) and oscillatory Alfv\'en waves.
}
\label{fig:thetaplot}
\end{figure}

One can investigate the effect of our free parameter $\theta$ on
the growth rate, and one finds that the growth rate can be greatly
enhanced, if $C^{(A-)}$ is very large compared to $C^{(s-)}$ and if
they are in phase, or almost stabilized if $C^{(A-)}$ is very large and
exactly out of phase with $C^{(s-)}$; but unless there is a significant
stabilizing gravitational acceleration, the flame remains unstable for
the entire parameter range of solutions.   A typical case is shown in
Fig.~\ref{fig:thetaplot}, with $\bar{a}_u = 1.3$, $\bar{g} = 0$, and
$\alpha = 2,3,4,5$.   Absent any selection criterion to select a unique
$\theta$, all that one can say is that the trans-alfv\'enic flames
are unstable, with essentially arbitrary growth rate.

% ============================================================================
%  field transverse to flame propagation
% ============================================================================
\section{FIELD TRANSVERSE TO DIRECTION OF FLAME PROPAGATION}
\label{sec:bperpendicular}
\subsection{Perturbation equations}
\label{sec:bperppert}

In the perpendicular-field case, the analysis is more straightforward as there
is no competition between the flame velocity and the Alfv\'en velocity.
We begin with Eqs.~\ref{eq:momentum}--\ref{eq:incompressible} from \S\ref{sec:bparallelpert}, 
but now our magnetic field is
\begin{eqnarray}
\label{eq:magpertsperp}
{\bf{\magH}} & = & (\magh_x+\magH, \magh_y, \magh_z), \quad \magh_x,\magh_y,\magh_z \ll \magH.
\end{eqnarray}

In this case, our linearized equations in terms of normal modes are
\begin{eqnarray}
\label{eq:xmomentummodeperp}
\nwd u                                                                   & = & -\frac{\imag k_x}{\rho} \delta p \\
\label{eq:ymomentummodeperp}
\nwd v - \frac{\magmu \magH}{4 \pi \rho} \left ( i k_x \magh_y - i k_y \magh_x \right ) & = & -\frac{\imag k_y}{\rho} \delta p \\
\label{eq:zmomentummodeperp}
\nwd w - \frac{\magmu \magH}{4 \pi \rho} \left ( i k_x \magh_z - D \magh_x \right )     & = & -\frac{1}{\rho} D \delta p \\
\label{eq:xinductionmodeperp}
\nwd \magh_x & = & i k_x \magH u \\
\label{eq:yinductionmodeperp}
\nwd \magh_y & = & i k_x \magH v \\
\label{eq:zinductionmodeperp}
\nwd \magh_z & = & i k_x \magH w \\
\label{eq:incompressiblemodeperp}
D w & = & - \left ( i k_x u + i k_y v \right ) \\
\label{eq:nomonopolesmodeperp}
D \magh_z & = & - \left ( i k_x \magh_x + i k_y \magh_y \right ).
\end{eqnarray}

Going through the same steps as in \S~\ref{sec:bparallel}, but
working with $\magh_x$ instead of $\magh_z$, we obtain:
\begin{eqnarray}
\label{eq:uperpdiffe}
\left ( \nwd^2 + (a k_x)^2 \right ) (D^2 - k^2) u & = & 0 \\
\label{eq:pperpdiffe}
\left ( \nwd^2 + (a k_x)^2 \right ) (D^2 - k^2) \delta p & = & 0 \\
\label{eq:hxperpdiffe}
\left ( \nwd^2 + (a k_x)^2 \right ) (D^2 - k^2) \nwd \magh_x & = & 0 .
\end{eqnarray}

Given the linearized equations, the perturbed quantities must have
$z$-dependencies which are linear combinations of the following:
\begin{equation}
\coefa e^{k z} + \coefb e^{-k z} + \coefc e^{-\frac{n}{W}z} + \coefd e^{-\frac{n + \imag a k_x}{W} z} + \coefe e^{-\frac{n - \imag a k_x}{W} z}.
\end{equation}

We can then find the expressions for the perturbed quantities
which satisfy the momentum and induction equations, and the divergence
criterion for the velocity and magnetic field:
\begin{eqnarray}
{\magh_x}_u & = & \magH_u \coefa_u e^{k z} \\
   u_u  & = & -\imag \frac{(n+W_u k)}{k_x} \coefa_u e^{k z} \\
\delta p_u & = & \frac{\rho_u (n + W_u k)^2}{k_x^2} \coefa_u e^{k z}\\
v_u & = & -\imag \frac{(n + W_u k) k_y}{k_x^2} \coefa_u e^{k z} \\
w_u & = & -\frac{(n + W_u k)\left(1 + \frac{k_y^2}{k_x^2}\right)}{k} \coefa_u e^{k z} \\
{\magh_y}_u & = & \magH_u \frac{k_y}{k_x} \coefa_u e^{k z} \\
{\magh_z}_u & = & -\imag \magH_u \frac{k}{k_x} \coefa_u e^{k z}
\end{eqnarray}

\begin{eqnarray}
{\magh_x}_b & = & \magH_b \coefb_b e^{-k z} +  \magH_b \coefd_b e^{-\frac{n + \imag a_b k_x}{W_b}z} + \magH_b \coefe_b e^{-\frac{n - \imag a_b k_x}{W_b}z}\\
u_b  & = & -\frac{\imag}{k_x} \left (n - W_b k \right)  \coefb_b e^{-k z} - 
      a_b \coefd_b e^{-\frac{n + \imag a_b k_x}{W_b}z} + 
      a_b \coefe_b e^{-\frac{n - \imag a_b k_x}{W_b}z}\\
{\delta p}_b  & = & 
    \rho_b \frac{(n - W_b k)^2}{k_x^2} \coefb_b e^{-k z} - \left (\rho_b a_b^2 \right ) \coefd_b e^{-\frac{n + \imag a_b k_x}{W_b}z} 
  - \left (\rho_b a_b^2 \right ) \coefe_b e^{-\frac{n - \imag a_b k_x}{W_b}z}\\
v_b  & = & -\imag k_y \frac{n - W_b k}{k_x^2 } \coefb_b e^{-k z} - 
            a_b \coefdp_b e^{-\frac{n + \imag a_b k_x}{W_b}z} + 
            a_b \coefep_b e^{-\frac{n - \imag a_b k_x}{W_b}z} \\
w_b  & = &  \frac{(n - W_b k)k}{k_x^2} \coefb_b e^{-k z} - 
      \imag \frac{W_b \left (k_x \coefd_b + k_y \coefdp_b \right ) }{n + \imag a_b k_x} a_b e^{-\frac{n + \imag a_b k_x}{W_b}z} + \nonumber \\
      & & \imag \frac{W_b \left (k_x \coefe_b + k_y \coefep_b \right )}{n - \imag a_b k_x} a_b  e^{-\frac{n - \imag a_b k_x}{W_b}z}\\
{\magh_y}_b  & = & 
    \frac{k_y}{k_x} \magH_b \coefb_b e^{-k z} + 
      \magH_b \coefdp_b e^{-\frac{n + \imag a_b k_x}{W_b}z} + \magH_b \coefep_b e^{-\frac{n - \imag a_b k_x}{W_b}z}\\
{\magh_z}_b  & = & \imag \magH_b \frac{k}{k_x} \coefb_b e^{-k z} + 
      \imag W_b \magH_b \frac{k_x \coefd_b + k_y \coefdp_b}{n + \imag a_b k_x} e^{-\frac{n + \imag a_b k_x}{W_b} z} + \nonumber \\
      & & \imag W_b \magH_b \frac{k_x \coefe_b + k_y \coefep_b}{n - \imag a_b k_x} e^{-\frac{n - \imag a_b k_x}{W_b} z}.
\end{eqnarray}
As in \S~\ref{sec:bparallel}, $\coefc_b$ must vanish.

\subsection{Jump Conditions}

The jump conditions will be the same as in \S\ref{sec:bparallelbcs},
but the components of $\bf{\magH}$ will
be different, and both tangential components must be considered since
one is now selected by the direction of the magnetic field.   We have
\begin{eqnarray}
{\bf{\magH_t}} & = & \left(\magH+\magh_x, \magh_y \right )\\
\magH_n & = & \magh_z - \frac{\imag k_x}{n} w \magH \\
{\bf{U_t}} & = & \left(u + \frac{i k_x}{n} w_i W, v + \frac{i k_y}{n} w_i W \right ) \\
U_n & = & W.
\end{eqnarray}

Then to zeroth order in the perturbed quantities, the jump conditions
give us
\begin{eqnarray}
\left [ \rho W \right ] & = & 0 \\
\rho W \left [ W \right ] + \left [ P + \frac{\magmu \magH^2}{8 \pi} \right ]& = & 0 \\
\left [ W \magH \right ] & = & 0,
\end{eqnarray}
and to linear order,
\begin{eqnarray}
\left [ \magh_z - \frac{i k_x}{n} w_i \magH \right ] & = & 0 \\
\rho W \left [ u + \frac{i k_x}{n} w_i W \right ] 
   - \frac{1 \magmu}{4 \pi} \left ( \magh_z - \frac{i k_x}{n} w_i \magH \right ) \left [ \magH \right ] & = & 0\\
\rho W \left [ v + \frac{i k_y}{n} w_i W \right ] & = & 0 \\
\left [ \delta p + \frac{\magmu \magH \magh_x}{4 \pi} \right ] & = & \frac{g w_i}{n} \left [ \rho \right ] \\
\left [ W \magh_x \right ] & = & 0 \\
\left [ W \magh_y \right ] & = & 0.
\end{eqnarray}

\subsection{Growth Rates}

\subsubsection{$k = k_y {\bf{\hat{y}}}$}

Note that by Eqs.~\ref{eq:uperpdiffe}, \ref{eq:pperpdiffe},
\ref{eq:hxperpdiffe} the magnetic field only enters through $a k_x$,
so that for $k_x = 0$ --- that is, the perturbed mode is orthogonal to
the magnetic field --- these equations reduce to the equations for the
Landau-Darrieus instability, and the magnetic field has no effect.

\subsubsection{$k = k_x {\bf{\hat{x}}}$}

In this case, as in \S~\ref{sec:bparallel}, again the
$\hat{\bf{y}}$-components decouple, and we are left with only the linear
equations for the $\hat{\bf{x}}$-components.  The linearized equations
have a non-zero solution only for
\begin{eqnarray}
%\left( 1 + {\alpha}^3 {c_u}^2 - 2 {\alpha}^2 {c_u}^2 \bar{n} + 
%      \alpha {c_u}^2 {\bar{n}}^2 \right)  \times \nonumber \\
%\left( -1 - \bar{n} + {\alpha}^3 {c_u}^2 \left( 1 + \bar{n} \right)  + 
%      \alpha \left( 2 + {c_u}^2 \left( 1 + \bar{n} \right)  \left( \bar{g} - {\bar{n}}^2 \right)  \right)  - 
%      {\alpha}^2 \left( 1 + \bar{n} \right)  \left( 1 + {c_u}^2 \left( \bar{g} + {\left( 1 + \bar{n} \right) }^2 \right) 
%         \right)  \right)  & = &0.
-\left( {\alpha}^3 + {{\bar{a}_u}}^2 - 2 {\alpha}^2 \bar{n} + \alpha {\bar{n}}^2 \right) \nonumber \\
    \left( {\alpha}^3 \left( 1 + \bar{n} \right)  - {{\bar{a}_u}}^2 \left( 1 + \bar{n} \right)  - 
      {\alpha}^2 \left( 1 + \bar{n} \right)  \left( {{\bar{a}_u}}^2 + \bar{g} + {\left( 1 + \bar{n} \right) }^2 \right)  + 
      \alpha \left( 2 {{\bar{a}_u}}^2 + \left( 1 + \bar{n} \right)  \left( \bar{g} - {\bar{n}}^2 \right)  \right)  \right)  & = & 0.
\end{eqnarray}
Factoring out trivial solutions $\bar{n} = (\alpha \pm \imag {\bar{a}_u} \alpha^{-1/2})$ corresponding to 
no wrinkling of the interface, we are left with
\begin{equation}
\label{eq:bperpgrowth}
\bar{n}^3 + 
\frac{1 + 3 \alpha}{1 + \alpha} \bar{n}^2 -
\frac{\alpha^3 - \alpha^2({\bar{a}_u}^2 + \bar{g} + 3) + \alpha \bar{g} -{\bar{a}_u}^2 }{\alpha (\alpha + 1)} \bar{n} -
\frac{(\alpha - 1) (\alpha^2 - \alpha({\bar{a}_u}^2 + \bar{g}) + {\bar{a}_u}^2) }{\alpha ( 1 + \alpha) } = 0
\end{equation}

In the limit of $\bar{a}_u \rightarrow 0$ this reduces to the correct
Landau-Darrieus limit.   Taking the other limit, in unscaled 
quantities, of $W_u \rightarrow 0$, one obtains
\begin{equation}
n^2 = g k \left ( \frac{\rho_b - \rho_u}{\rho_b + \rho_u} - \frac{B_u^2 k^2 \left ( 1 + \alpha^2 \right )}{4 \pi g k^3 \left ( \rho_b + \rho_u\right )}  \right ).
\label{eq:magrt}
\end{equation}
Note that the second term in this expression is different than the result
quoted in \cite{chandra}~\S~97, where it is assumed that the horizontal
field is equal in both regions of the domain, and so we have here $B_u^2
(1 + \alpha^2) = B_u^2 + B_b^2$ rather than $2 B^2$.   If one repeats
the derivation provided therein with $\alpha B_b = B_u$, as we have here,
then the two results agree.

\begin{figure}
\plotone{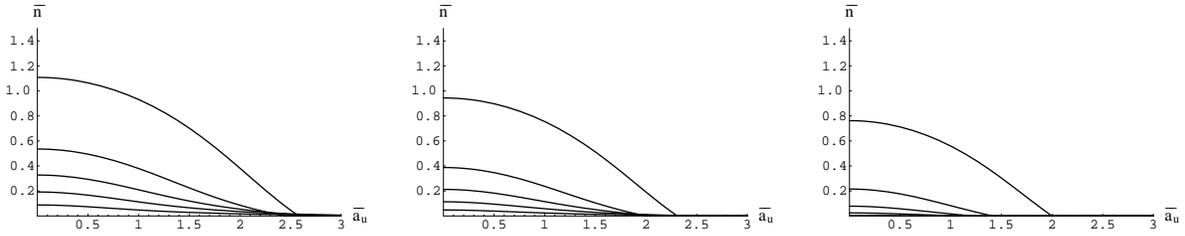}
\caption{The maximum real part of the scaled growth rate ($\bar{n}$) for the case of a perpendicular
magnetic field as a function of $\bar{a}_u$, with
$\bar{g} = -1$ on the left, $\bar{g} = 0$ in the center, and $\bar{g} =
+1$ on the right.  Lines are plotted for, top to bottom, $\alpha$ = 4, 2, 1.5, 1.25, and 1.1}
\label{fig:perpendicular-growthvsalpha}
\end{figure}

\begin{figure}
\plotone{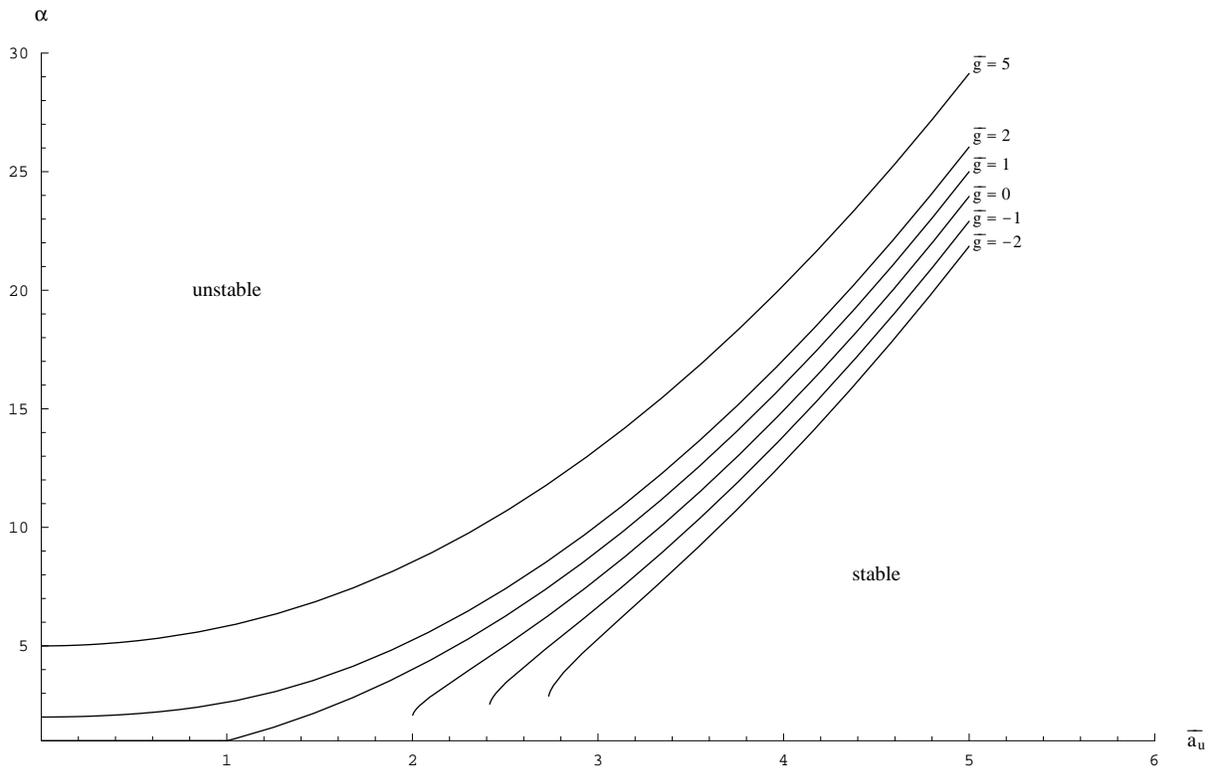}
\caption{Stability boundaries in $\alpha$ for the case of a perpendicular
magnetic field as a function of $\bar{a}_u$, for (top to bottom) $\bar{g}$ = 5,2,1,0,-1,-2.}
\label{fig:perpendicular-stability}
\end{figure}

The maximum real solution of Eq.~(\ref{eq:bperpgrowth}) is plotted in
Fig.~\ref{fig:perpendicular-growthvsalpha}.  We see that for $\bar{a}_u>1$
there is significant suppression of growth of the instability, 
One can examine the stability boundary by considering $\bar{n} = 0$
in Eq.~(\ref{eq:bperpgrowth}).   This gives us
\begin{equation}
\alpha^2 -  (\bar{a}_u^2 + \bar{g}) \alpha + \bar{a}_u^2 = 0.
\end{equation}
The minimum $\alpha$ for instability as a function of $\bar{a}_u$, for
varying $\bar{g}$, is plotted in Fig.~\ref{fig:perpendicular-stability}.
Since a typical value for $\alpha$ for a carbon flame in a degenerate white dwarf
is $\approx 1.3-2.4$ (\eg, \citealt{flamecurvature}), one sees
that once $\bar{a}_u$ exceeds 1, flame instability is greatly suppressed.

It should be noted here that the density ratio $\alpha$, which in the
non-magnetic case or in the case of \S~\ref{sec:bparallel} is given by
the physicochemical properties of the burning and the EOS of the fluid,
here is also in principle a function of the magnetic field, as the ambient
field also provides pressure support.  However, in the limit where the sound
speed is much larger than the Alfv\'en speed, this contribution is small.

% ============================================================================
%  Local effects aside from instability
% ============================================================================
\section{LOCAL EFFECTS OF MAGNETIC FIELD ON FLAME VELOCITY}
\label{sec:localspeed}
Astrophysical flames propagate by thermal diffusion.  In the deep interior
of a white dwarf, thermal diffusion is due largely to electrons; in
the presence of a magnetic field, this diffusion will be anisotropic.
(In outer regions of the star, where thermal diffusion is due to
photons, there will be no such effect.)  One must then investigate the
range of validity of the assumption used in \S~\ref{sec:bparallel} and
\S~\ref{sec:bperpendicular} that there will be a constant flame speed
despite the wrinkling (and thus differing orientations) of the flame.

We will consider here a very simple model for the anisotropic thermal diffusion,
\begin{equation}
D_\mathrm{th}(\theta) = D_\mathrm{th}(0) \frac{a + b |\cos \theta|}{a + b}
\end{equation}
where $D_\mathrm{th}$ is the thermal diffusivity, $\theta$ represents
the angle between the direction of diffusion --- that is, the local direction
of flame propagation --- and the ambient magnetic field, and $a$ and $b$
parameterize the anisotropy.

The local flame normal, from \S~\ref{sec:bparallel} (but not restricted
to linear order) is given by
\begin{equation}
{\bf{\hat{n}}} = \left ( \frac{\partial}{\partial x} z_f, \sqrt{1 - \left ( \frac{\partial}{\partial x} z_f \right )^2} \right ) = \left ( -\epsilon, \sqrt{1 - \epsilon^2} \right ).
\end{equation}

Simple expressions for flame velocity (\eg, \citealt{flamespeed}) find that the
propagation velocity scales with the square root of the diffusivity.  
Thus, we will consider the angle dependence of the flame velocity
\begin{equation}
S_l (\theta) = S_l(0) \sqrt{\frac{D_\mathrm{th}(\theta)}{D_\mathrm{th}(0)}} = S_l(0) \sqrt{\frac{a + b | \cos \theta |}{a + b}}.
\end{equation}

\subsection{Field parallel to flame propagation}

In the case of the ambient magnetic field parallel to the direction of
flame propagation discussed in \S~\ref{sec:bparallel}, with ${\bf{B}} =
B {\bf{\hat{z}}}$.   We then find
\begin{equation}
\cos \theta = {\bf{\hat{B}}}\cdot{\bf{\hat{n}}} = \sqrt{1 - k^2 z_f^2} = \sqrt{1 - \epsilon^2},
\end{equation}
where $z_f k = \epsilon \ll 1$ as in Appendix~\ref{sec:app:linearization}.

If we then expand our expression for $S_l(\theta)$ in terms of the
small quantity $k z_f$ one finds
\begin{equation}
\frac{S_l(\theta)}{S_l(0)} = 
1 - \frac{b}{4 \left( a + b \right) }  \epsilon  ^2  + O\left( \epsilon^4 \right ),
\end{equation}
that is, there is no correction to the flame speed to the linear order
of the analysis we have performed in \S~\ref{sec:bparallel}.  
Thus, it is reasonable to consider a constant flame velocity in this
case, at least for the simple angular dependence considered here.

\subsection{Field perpendicular to flame propagation}

On the other hand, for the case of the flame propagating across
the magnetic field considered in \S~\ref{sec:bperpendicular},
${\bf{B}} = B{\bf{\hat{x}}}$.  In this case, $\cos \theta =
{\bf{\hat{B}}}\cdot{\bf{\hat{n}}}  = {\bf{\hat{x}}}\cdot{\bf{\hat{n}}}  = - \epsilon $ which is of linear
order, not quadratic order, in small quantities.   In this case, we find
\begin{equation}
\frac{S_l(\theta)}{S_l(\pi/2)} = 1 + \frac{b}{2 a} \epsilon + O\left ( \epsilon^{2}\right )
\end{equation}
so that for consistency to linear order, in this case the thermal
diffusion must be nearly isotropic ($a \gg b$ --- in particular, $b/a$ must
be of order $\epsilon$).   Note that for a perpendicular
field, there would have to be a significant isotropic component to the
diffusion anyway for there to be a meaningful propagating flame, since
the flame must propagate by diffusion across the magnetic field;
however, this is a weaker condition than $b/a \sim \epsilon$.   Note
that for diffusivities which depend more strongly on $\theta$, such
as $\cos^2(\theta)$, the flame speed will no longer depend on linear
orders of perturbed quantities even for the perpendicular field.

% ============================================================================
%  Markstein-length effect on flame
% ============================================================================
\section{EFFECTS OF CURVATURE}
\label{sec:markstein}
Independent of the effects of a magnetic field, it is well known that
the velocity of a laminar flame depends on the details of the flame
structure, and that curvature of the flame will change the flame speed.
A simple relation for the modified flame speed was first given by
\cite{markstein64},
\begin{equation}
S_l = S_l^0 \left ( 1 + l_M \frac{\partial^2 z_f}{{\partial x}^2} \right )
\end{equation}
where $S_l^0$ is the planar flame speed, $S_l$ is the modified
flame speed, and $l_M$ is an empirically determined `Markstein
length', which is divided by the radius of curvature of the flame.
This simple prescription has proved very robust, and Markstein lengths
for some relevant astrophysical flames were numerically measured in
\cite{flamecurvature}.

This non-constancy of the flame speed can be incorporated into the
results of \S\ref{sec:bparallel} and \S\ref{sec:bperpendicular}.
One first observes that in the unperturbed frame, the flame now
moves with a velocity
\begin{equation}
W + w_i - S_l = w_i + k^2 z_f l_M W \quad , 
\end{equation} 
and thus the perturbed flame position $z_f$ is now 
\begin{eqnarray}
n z_f & = & w_i + k^2 z_f l_M W\\
  z_f & = & \frac{w_i}{n - k^2 l_M W}.
\label{eq:newzf}
\end{eqnarray}
Because the flame position must be the same in both regions of the
flame, the jump condition Eq.~(\ref{eq:noholes}) becomes
\begin{equation}
\left [ w_i \right ] + k^2 z_f l_M \left [ W \right ] = 0.
\end{equation}

In the case of the magnetic field parallel to the direction of propagation
of the flame, jump conditions implicitly involving $z_f$ must be modified
appropriately to use Eq.~(\ref{eq:newzf}), and Eq.~(\ref{eq:momncontin})
has a new term directly from the modified flame speed.   The jump
conditions become
\begin{eqnarray}
\left [ \magh_z \right ] & = & 0 \\
\rho W \left [ u + i k z_f W \right ] - \frac{\magmu \magH}{4 \pi} \left [ \magh_x \right ] & = & 0 \\
\left [ \delta p \right ] - 2 k^2 z_f l_M \rho W \left [ W \right ] & = & g z_f \left [ \rho \right ] \\
\left [ W \magh_x \right ] - \magH \left [ u \right ] & = & 0.
\end{eqnarray}
Again, the final jump condition may be discarded.

One can apply these boundary conditions and find the solution of the linear equations
as in \S\ref{sec:bparallel}.   For super-Alfv\'enic flames, one obtains for the growth rate
\begin{eqnarray}
  {\bar{n}}^3  
  + \frac{\left (1 + 3 \alpha\right ) - 2 \alpha \bar{l}_M }{1 + \alpha} \bar{n}^2 - 
  \frac{\left(\alpha^2 + \alpha \left( 2 {\bar{a}_u}^2 - \bar{g} - 3\right) + \bar{g}\right) + 2 \alpha \left ( \alpha + 1 \right ) \bar{l}_M }{1 + \alpha} \bar{n} \nonumber \\ 
- \frac{\left (\alpha - 1 \right ) \left( \alpha - \bar{g} \right) + 2 \alpha \bar{l}_M \left ( \alpha - {\bar{a}_u}^2\right)}{1 + \alpha} & = & 0,
\end{eqnarray}
where $\bar{l}_M = k l_M$ is the non-dimensionalized Markstein length.   Note that the above expression
reduces to Eq.~(\ref{eq:bparsupergrowth}) for $\bar{l}_M \rightarrow 0$.

\begin{figure}
\plotone{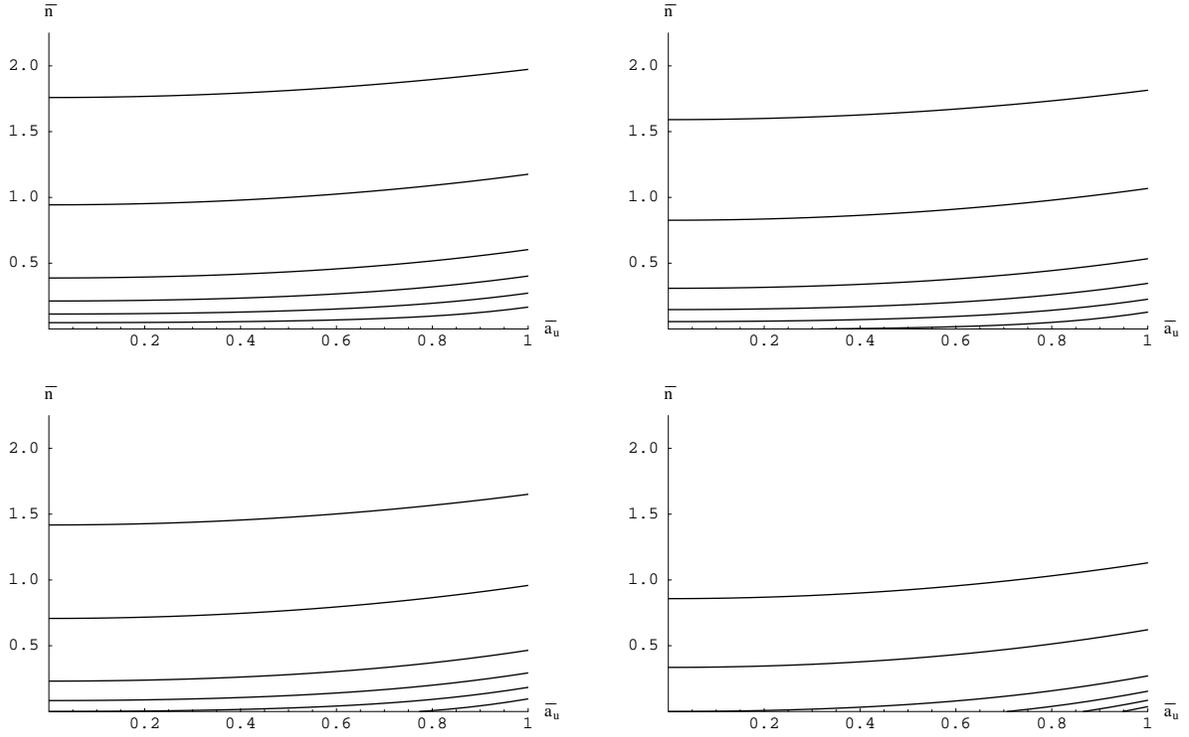}
\caption{The growth rate for a parallel magnetic field when the flame
is everywhere super-Alfv\'enic, and has a negative Markstein length.
We plot the maximum real part of the scaled growth rate ($\bar{n}$)
as a function of $\bar{a}_u$, with $\bar{l}_M = 0$ on
the top left, $\bar{l}_M = -0.05$ on top right, $\bar{l}_M = -0.10$
on the bottom left, and $\bar{l}_M = -0.25$ on the bottom right.
Lines are plotted for, top to bottom, $\alpha$ = 8, 4, 2, 1.5, 1.25, 1.1.
}
\label{fig:parallel-super-markstein}
\end{figure}

We can see how the Markstein length effects the growth rate in Fig.~\ref{fig:parallel-super-markstein}.
If the Markstein length is comparable to the wavelength of the perturbation ($k l_M \approx 1$), the
instability is greatly suppressed.    

Since the effect for other cases is similar, we don't show growth rate plots, but for completeness,
we include the expressions for growth rate here.  For sub-Alfv\'enic flames, using the same jump conditions
one obtains
\begin{eqnarray}
{\bar{n}}^3  - \frac{{\alpha}^{\frac{3}{2}} - {\sqrt{\alpha}} \left( 1 + 2 \bar{a}_u \right)  - 
      2 \alpha \bar{a}_u + 2 \alpha \bar{l}_M }{1 + \alpha} \bar{n}^2 \nonumber \\
\frac{{\alpha}^2 - 2 {\alpha}^{\frac{3}{2}} \left( 2 + \bar{a}_u \right)  - \bar{g} + 
    \alpha  \left( 1 + 2 \bar{a}_u + 2 \bar{a}_u^2 + \bar{g} \right) }{1 + \alpha} \bar{n} 
 -\bar{l}_M \frac{2 \alpha \bar{a}_u \left ( 1 + \sqrt{\alpha} \right )}{1 + \alpha}  \bar{n} \nonumber \\
\frac{\left( -1 + {\sqrt{\alpha}} \right)  {\sqrt{\alpha}} 
    \left( \alpha + {\alpha}^{\frac{3}{2}} + \bar{g} - {\sqrt{\alpha}} \bar{g} + 2 \bar{a}_u \bar{g} \right) }{1 + 
    \alpha} + \nonumber \\
\bar{l}_M \frac{2 \alpha^{3/2} \left ( 1 + \bar{a}_u - \bar{a}_u^2 - \sqrt{\alpha} \left ( 1 + \bar{a}_u \right ) + \alpha \right ) }{1 + \alpha} & = & 0.
\end{eqnarray}

Trans-Alfv\'enic flames remain non-evolutionary, but again parameterizing the range of possible solutions
one obtains
\begin{eqnarray}
{\bar{n}}^3  
+ \frac{\left(4 \alpha - \theta\right) - 2 \alpha \bar{l}_M }{1 + \alpha} \bar{n}^2 \nonumber \\
- \frac{\left ( \alpha^2 + \alpha \left ( \bar{a}_u^2 - \bar{g} - 3 \right) + \bar{g} \right) + 2 \alpha \left ( 1 + \alpha \right ) \bar{l}_M}{1 + {\alpha}} \bar{n} \nonumber \\
+ \frac{\theta (\bar{g}-\alpha) + 2 \alpha \bar{l}_M \left ( \bar{a}_u^2 - \theta - 1 \right )} {1 + \alpha} & = & 0
\end{eqnarray}

In the case of a magnetic field parallel to the flame, as in \S\ref{sec:bperpendicular}, 
the jump conditions become
\begin{eqnarray}
\left [ \magh_z - i k_x z_f \magH \right ] & = & 0 \\
\rho W \left [ u + i k_x z_f W \right ] 
   - \frac{1 \magmu}{4 \pi} \left ( \magh_z - i k_x z_f  \magH \right ) \left [ \magH \right ] & = & 0\\
\rho W \left [ v + i k_y z_f  W \right ] & = & 0 \\
\left [ \delta p + \frac{\magmu \magH \magh_x}{4 \pi} \right ] - 2 k^2 z_f l_M \rho W \left [ W \right ]& = & g z_f \left [ \rho \right ] \\
\left [ W \magh_x \right ] & = & 0 \\
\left [ W \magh_y \right ] & = & 0.
\end{eqnarray}

The modified expression for growth rate then becomes
\begin{eqnarray}
{\bar{n}}^3  
+ \frac{1 + 3 \alpha - 2 \alpha \bar{l}_M}{1 + \alpha} \bar{n}^2 
+ \frac{-\alpha^3 + {\alpha}^2 \left( 3 + {\bar{a}_u}^2 + \bar{g} \right)  
- \alpha \bar{g} + \bar{a}_u^2 - 2 \bar{l}_M \alpha^2 \left(\alpha + 1\right)}{\alpha \left( 1 + \alpha \right)}
\bar{n} \nonumber \\
- \frac{
\left ( \alpha - 1 \right ) 
\left ( \alpha^2 - \alpha \left ( \bar{a}_u^2 + \bar{g}^2 \right ) + \bar{a}_u^2 \right )
+ 2 \alpha \bar{l}_M \left ( \alpha^2 + \bar{a}_u^2 \right )
}{\alpha \left( 1 + \alpha \right)} 	& = & 0.
\end{eqnarray}

% ============================================================================
%  Conclusions
% ============================================================================
\section{CONCLUSIONS}
\label{sec:conclusions}
We have examined the growth of a small perturbation to a flame in ideal
MHD, and compared the growth to the well-known  results in the absence
of a field.   We find that for the magnetic field to greatly suppress
the flame instability requires $a_u > W_u$.   This requires extremely
strong magnetic fields.  Assuming some of the largest-observed magnetic
fields in white dwarfs, we have
\begin{equation}
a_u \sim 3 \times 10^4 \cms \left ( \frac{B}{10^9 G} \right ) \left ( \frac{\rho}{10^8 \gcc} \right )^{-1/2}.
\end{equation}
With carbon-oxygen flame speeds in the core of a white dwarf more on
the order of $10^6 \cms$ \citep{timmeswoosley}, this is likely too low to
play a major in the entire star.    However, localized strong magnetic field
regions could easily have an effect; and it is known \citep{cattaneo99}
that in chaotic flows such as the turbulence that is likely to exist in
a rotating white dwarf that even with originally extremely weak magnetic
fields, local peaks are quickly generated with $\bar{a}_u \approx 1$.

On the surface of a neutron star, where the magnetic field might be the
same but densities would be lower,
\begin{equation}
a_u \sim 1 \times 10^6 \cms \left ( \frac{B}{10^9 G} \right ) \left (
\frac{\rho}{10^5 \gcc} \right )^{-1/2},
\end{equation} 
and magnetic field effects might well be significant everywhere.

Weaker magnetic fields also effect flame growth, but only modestly.
If the flame is everywhere super-Alfv\'enic, the effects of magnetic field
will somewhat stabilize the flame if the magnetic field is perpendicular
to the flame propagation, and destabilize it if it is parallel.

It is instructive to compare the results presented here with those known
for MHD shocks, where fast parallel shocks (which are super-Alfv\'enic)
and perpendicular shocks are stable under a broad range of parameters
\citep{gardnerkruskal}, while slow shocks (which are sub-Alfv\'enic)
are broadly unstable \citep{lesson,edelman90b,stoneedelman}; this is
just the opposite of what is reported here.  The key difference between
the two, for the purposes of this instability, is the direction of the
density jump across the discontinuity --- recall that it is this density
jump which drives the instability.

In the flames discussed here, heating from the reactions ensures that
$\rho_b < \rho_u$; that is, $\alpha > 1$.  However, in the case of a
shock, in our notation $\alpha < 1$.  This changes the underlying driving
of the instability; indeed, if one uses $\alpha < 1$ in the growth rates
presented above, one finds super-Alfv\'enic and parallel flames to be
stable, and sub-Alfv\'enic flames to be unstable.

This difference also changes the properties of the nonevolutionary
nature of the interface in the trans-Alfv\'enic case.   In the case of
a trans-Alfv\'enic MHD shock, there is one too few Alfv\'en waves to
satisfy the boundary conditions, so there is no solution; in the case
of a trans-Alfv\'enic flame, there is one too many, so there are an
infinite number of solutions, and no unique solution.   This will change
the nature of the behavior in this regime.  Future work must include
numerical simulations of the interesting trans-Alfv\'enic flame regime,
which could occur locally in a white dwarf and in large regions on the
surface of a neutron star.

\acknowledgements 
            This work was supported by the Department of Energy
Computational Science Graduate Fellowship Program of the Office of
Scientific Computing and Office of Defense Programs in the Department of
Energy under contract DE-FG02-97ER25308.   The author thanks R. Rosner,
A. Calder, and T. Dupont for helpful comments, and R. Morgan-Dursi for her support.

\appendix
\section{Linearization}
\label{sec:app:linearization}
\subsection{MHD equations}
\label{sec:app:linmhd}

In this section, we describe in more detail our process of linearizing
the ideal MHD equations.

We perturb the initially planar flame with a very small amplitude
wrinkle, giving it an initial perturbed position $z_f$.    We require
the amplitude to be small compared to the wavelength of the perturbation:

\begin{equation}
\frac{2 \pi z_f}{\lambda} = z_f k = \epsilon \ll 1.
\end{equation}

Such a perturbation will generate perturbations in the magnetic
fields and velocities of the same order; we expand them in series of this perturbation
parameter $\epsilon$.   In the case of a parallel magnetic field,
\begin{eqnarray}
{\bf{B}} & = & B \left ( \epsilon {b_x}_1 + \epsilon^2 {b_x}_2 + \cdots, 
 \epsilon {b_y}_1 + \epsilon^2 {b_y}_2 + \cdots, 
 1 + \epsilon {b_z}_1 + \epsilon^2 {b_z}_2 + \cdots \right ) \\
{\bf{U}} & = & W \left ( \epsilon {u}_1 + \epsilon^2 {u}_2 + \cdots, 
 \epsilon {v}_1 + \epsilon^2 {v}_2 + \cdots, 
 1 + \epsilon {w}_1 + \epsilon^2 {w}_2 + \cdots \right ) .
\end{eqnarray}

Because the flow is incompressible, the pressure fluctuations will be fluctuations in 
the dynamic, not thermodynamic, pressure; thus the pressure fluctuations scale with
$\rho W^2$ rather than $P$:
\begin{equation}
p = P + \rho W^2 \epsilon \delta p_1 + \rho W^2 \epsilon^2 \delta p_2 + \cdots.
\end{equation}

We assume that all the non-dimensional expansion coefficients are of
order unity.  We will also assume that the non-dimensional parameters of
the problem --- the scaled growth rate ($\bar{n} = n/(k W))$, gravity
($\bar{g} = g/(k W^2)$) and Alfv\'en speed ($\bar{a}_u = a_u/W_u$)
--- are order unity quantities.

We may now write the ideal MHD equations in terms of these expansions.   $W$, $B$, and $\epsilon$
are independent of time or position, and we retain terms up to first order in $\epsilon$.  For the momentum equation, we then have:
\begin{eqnarray}
\epsilon \frac{\partial u_1}{\partial t} + W \epsilon \frac{\partial u_1}{\partial z} - \frac{B^2 \epsilon}{4 \pi \rho W} \left ( \frac{\partial {b_x}_1}{\partial z} - \frac{\partial {b_z}_1}{\partial x} \right ) & = & - W \epsilon \frac{\partial \delta p_1}{\partial x} \\
\epsilon \frac{\partial v_1}{\partial t} + W \epsilon \frac{\partial v_1}{\partial z} - \frac{B^2 \epsilon}{4 \pi \rho W} \left ( \frac{\partial {b_y}_1}{\partial z} - \frac{\partial {b_z}_1}{\partial y} \right ) & = & - W \epsilon \frac{\partial \delta p_1}{\partial y} \\
\epsilon \frac{\partial w_1}{\partial t} + W \epsilon \frac{\partial w_1}{\partial z} & = & - W \epsilon \frac{\partial \delta p_1}{\partial z}.
\end{eqnarray}

Re-expressing the linear-order perturbed quantities in dimensional units, \eg{}
$(u,v,w) = \epsilon W (u_1, v_1, w_1)$, $(b_x, b_y, b_z) = \epsilon B
({b_x}_1,{b_y}_1, {b_z}_1)$, $\delta p = \epsilon \rho W^2 \delta p_1$,
we obtain Eqns.~(\ref{eq:xmomentumpert}--\ref{eq:zmomentumpert}).
Other equations, and the boundary conditions, are linearized in the
same manner.

\subsection{Matching Boundary Conditions}
\label{sec:app:linbcs}

In applying the boundary conditions at $z = z_f$, we can consider
expanding the exponential $z$-dependencies in Taylor series.  In the
case of the magnetic field, expanding the exponential terms 
on the unburned side (the terms on the burned side are analogous)
one obtains:
\begin{eqnarray}
e^{-k z_f} & = & 1 - \epsilon + \epsilon^2 + \cdots \\
e^{-\frac{n}{W_u + a_u} z_f} & = & 1 - \frac{\bar{n}}{1+\bar{a}_u} \epsilon +\frac{\bar{n}^2}{(1+\bar{a}_u)^2} \epsilon^2 +  \cdots \\
\label{eq:camterm}
e^{-\frac{n}{W_u - a_u} z_f} & = & 1 - \frac{\bar{n}}{1-\bar{a}_u} \epsilon +\frac{\bar{n}^2}{(1-\bar{a}_u)^2} \epsilon^2 +  \cdots,
\end{eqnarray}
while the values for at $z = 0$ would clearly be exactly unity.  Since
these terms multiply the already order-$\epsilon$ coefficients $C^{(i)}$,
then, the difference in the perturbed quantities between their values at
$z = 0$ and $z = z_f$ is of order $\epsilon^2$ and can be neglected in
this analysis as long as $1 - \bar{a}_u \not \approx \epsilon$.    As we
will see, however, as $\bar{a}_u \rightarrow 1$, the terms of the form
(\ref{eq:camterm}) vanish, removing this restriction, and we may consider 
$z_f = 0$ for the boundary conditions.

\clearpage
\bibliographystyle{plainnat}
\bibliography{ldmhd}

% ============================================================================
%  Figures
% ============================================================================
\clearpage

%%%%%%%%%%%
%% Introduction figures

\clearpage
\newpage

%%%%%%%%%%%
%% Parallel figures

% subalfenic 

% stability boundaries

\clearpage
\newpage

% bt jump

\clearpage
\newpage
%%%%%%%%%%%
%% Perpendicular figures

\clearpage
\newpage

%%%%%%%%%%%
%% Perpendicular figures

%%%%%%%%%%%%
%% Markstein figures

\end{document}